\documentclass[twocolumn]{aastex701} 

\newcommand{\citefull}[1]{\citeauthor{#1} \citeyear{#1}}
\usepackage{makecell}
\usepackage{tablefootnote}
\usepackage{threeparttable}
\usepackage{amsmath}

\shorttitle{Cold Giants from RVs and Astrometry}
\shortauthors{Wu et al.}
\begin{document}

\title{Long-period Giant Companions to HD 361, HD 196068, and HD 203432: Joint RV and Astrometric Constraints\footnote{This paper includes data gathered with the 6.5 meter Magellan Telescopes located at Las Campanas Observatory, Chile.}}

\author[0009-0005-1518-8258]{Yiyun Wu}
\affiliation{State Key Laboratory of Dark Matter Physics, Tsung-Dao Lee Institute \& School of Physics and Astronomy, Shanghai Jiao Tong University, Shanghai 201210, China}
\email{yy{\_}cloud@sjtu.edu.cn}

\author[0009-0000-9341-3442]{Yifan Xuan}
\affiliation{State Key Laboratory of Dark Matter Physics, Tsung-Dao Lee Institute \& School of Physics and Astronomy, Shanghai Jiao Tong University, Shanghai 201210, China}
\email{yifan.xuan@sjtu.edu.cn}  

\author[0000-0001-6753-4611]{Guang-Yao Xiao}
\affiliation{State Key Laboratory of Dark Matter Physics, Tsung-Dao Lee Institute \& School of Physics and Astronomy, Shanghai Jiao Tong University, Shanghai 201210, China}
\email{gyxiao{\_}tdli@sjtu.edu.cn}

\author[0000-0003-1305-3761]{R. Paul Butler}
\affiliation{Earth and Planets Laboratory, Carnegie Institution for Science, 5241 Broad Branch Road, NW, Washington, DC 20015, USA}
\email{pbutler@carnegiescience.edu}

\correspondingauthor{Fabo Feng}
\author[0000-0001-6039-0555]{Fabo Feng}
\affiliation{State Key Laboratory of Dark Matter Physics, Tsung-Dao Lee Institute \& School of Physics and Astronomy, Shanghai Jiao Tong University, Shanghai 201210, China}
\email[show]{ffeng@sjtu.edu.cn}

\author[0000-0001-5695-8734]{Zitao Lin}
\affiliation{Department of Astronomy, Tsinghua University, Beijing 100084, China}
\email{lzt22@mails.tsinghua.edu.cn}

\author[0009-0002-5981-8520]{Jiayue Zhang}
\affiliation{Department of Astronomy, Tsinghua University, Beijing 100084, China}
\email{zhangjy23@mails.tsinghua.edu.cn}

\author[0009-0003-8853-4540]{Jiayin Li}
\affiliation{Department of Astronomy, Tsinghua University, Beijing 100084, China}
\email{lijiayin416@gmail.com}

\author[0009-0008-8931-1513]{Bryant Randolph}
\affiliation{State Key Laboratory of Dark Matter Physics, Tsung-Dao Lee Institute \& School of Physics and Astronomy, Shanghai Jiao Tong University, Shanghai 201210, China}
\email{brandolph@sjtu.edu.cn}

\author[0009-0005-4888-2158]{Gaoyang Du}
\affiliation{Dalian University of Technology}
\email{gaoyangdu311@gmail.com}

\author[0009-0008-2801-5040]{Johanna K. Teske}
\affiliation{Earth and Planets Laboratory, Carnegie Institution for Science, 5241 Broad Branch Road, NW, Washington, DC 20015, USA}
\email{jteske@carnegiescience.edu}

\author[0000-0002-8681-6136]{Stephen A. Shectman}
\affiliation{Observatories of the Carnegie Institution for Science, 813 Santa Barbara Street, Pasadena, CA 91101, USA}
\email{shec@carnegiescience.edu}

\author[0000-0002-5226-787X]{Jeffrey D. Crane}
\affiliation{Observatories of the Carnegie Institution for Science, 813 Santa Barbara Street, Pasadena, CA 91101, USA}
\email{crane@carnegiescience.edu}

\author[0000-0003-0433-3665]{Hugh R. A. Jones}
\affiliation{Centre for Astrophysics Research, University of Hertfordshire, College Lane, AL10 9AB, Hatfield, UK}
\email{h.r.a.jones@herts.ac.uk}

\author[0000-0002-7595-0970]{C. G. Tinney}
\affiliation{School of Physics and Australian Centre for Astrobiology, University of New South Wales, Sydney 2052, Australia}
\email{c.tinney@unsw.edu.au}

\author[0000-0001-9957-9304]{Robert A. Wittenmyer}
\affiliation{Centre for Astrophysics, University of Southern Queensland, Toowoomba, QLD 4350, Australia}
\email{rob.wittenmyer@usq.edu.au}

\author[0000-0002-5726-7000]{Jeremy Bailey}
\affiliation{School of Physics and Australian Centre for Astrobiology, University of New South Wales, Sydney, NSW 2052, Australia}
\email{j.bailey@unsw.edu.au}

\author[0000-0003-2839-8527]{Simon J. O'Toole}
\affiliation{Australian Astronomical Optics, Macquarie University, North Ryde, NSW 1670, Australia}
\email{simon.otoole@mq.edu.au}

\author[0000-0003-0035-8769]{Brad D. Carter}
\affiliation{Centre for Astrophysics, University of Southern Queensland, Toowoomba, QLD 4350, Australia}
\email{carterb@usq.edu.au}

\author[0000-0002-6937-9034]{Sharon Xuesong Wang}
\affiliation{Department of Astronomy, Tsinghua University, Beijing 100084, China}
\email{sharonw@tsinghua.edu.cn}

% PFS observers appended in Johanna Teske's supplied order, 2026-11-12.
% Coauthor invitations/acceptance and preferred affiliations remain to be confirmed.

\author[0000-0001-8362-3462]{Mauro Barbieri}
\affiliation{Instituto de Astronom\'ia y Ciencias Planetarias, Universidad de Atacama, Copiap\'o, Chile}
\email{maurobarbieri.uda.cl@gmail.com}

% ORCID for Yuri Beletsky: NEEDS CONFIRMATION; no verified public match found.

\author{Yuri Beletsky}
\affiliation{Las Campanas Observatory, Carnegie Institution for Science, La Serena, Chile}
\email{ybeletsky@carnegiescience.edu}

\author[0000-0002-0040-6815]{Jennifer A. Burt}
\affiliation{Jet Propulsion Laboratory, California Institute of Technology, Pasadena, CA 91109, USA}
\email{jennifer.burt@jpl.nasa.gov}

\author[0000-0002-6939-9211]{Tansu Daylan}
\affiliation{Department of Physics, Washington University in St. Louis, St. Louis, MO 63130, USA}
\email{tansu@wustl.edu}

\author[0000-0002-2100-3257]{Mat\'ias R. D\'iaz}
\affiliation{Las Campanas Observatory, Carnegie Institution for Science, La Serena, Chile}
\email{mdiaz@carnegiescience.edu}

\author[0009-0006-4294-6760]{Adam Distler}
\affiliation{Center for Astrophysics | Harvard \& Smithsonian, 60 Garden Street, Cambridge, MA 02138, USA}
\email{adam.distler@cfa.harvard.edu}

\author[0000-0003-2313-467X]{Diana Dragomir}
\affiliation{Department of Physics and Astronomy, University of New Mexico, 210 Yale Blvd NE, Albuquerque, NM 87106, USA}
\email{dragomir@unm.edu}

\author[0000-0002-2482-0180]{Zahra Essack}
\affiliation{Department of Physics and Astronomy, University of New Mexico, 210 Yale Blvd NE, Albuquerque, NM 87106, USA}
\email{zessack@unm.edu}

\author[0000-0002-9464-8101]{Adina D. Feinstein}
\affiliation{Department of Physics and Astronomy, Michigan State University, East Lansing, MI, USA}
\email{adina@msu.edu}

\author[0000-0001-8045-1765]{Erin E. Flowers}
\affiliation{Center on Science and Technology, Princeton University, Princeton, NJ, USA}
\email{eflowers@princeton.edu}

\author[0000-0002-2036-2311]{Yadira S. Gaibor}
\affiliation{MIT Kavli Institute for Astrophysics and Space Research, Massachusetts Institute of Technology, 70 Vassar Street, Cambridge, MA 02139, USA}
\email{ygaibor@mit.edu}

\author[0000-0001-9261-8366]{Jhon Yana Galarza}
\affiliation{Departamento de Astronom\'ia, Universidad de Concepci\'on, Concepci\'on, Chile}
\email{jyanagalarza@carnegiescience.edu}

%\author[0000-0001-8401-4300]{Shubham Kanodia}
%\affiliation{Earth and Planets Laboratory, Carnegie Institution for Science, 5241 Broad Branch Road, NW, Washington, DC 20015, USA}
%\email{skanodia@carnegiescience.edu}

% ORCID for Adriana Kuehnel: NEEDS CONFIRMATION; no verified public match found.

\author{Adriana Kuehnel}
\affiliation{Earth and Planets Laboratory, Carnegie Institution for Science, 5241 Broad Branch Road, NW, Washington, DC 20015, USA}
\email{adrianakuehnel@hotmail.com}

\author[0000-0001-9269-8060]{Michelle Kunimoto}
\affiliation{Department of Physics and Astronomy, University of British Columbia, Vancouver, BC, Canada}
\email{michelle.kunimoto@gmail.com}

\author[0000-0001-7516-8308]{Benjamin T. Montet}
\affiliation{School of Physics, University of New South Wales, Sydney, NSW 2052, Australia}
\email{b.montet@unsw.edu.au}

\author[0000-0003-0412-9664]{David J. Osip}
\affiliation{Las Campanas Observatory, Carnegie Institution for Science, La Serena, Chile}
\email{dosip@carnegiescience.edu}

\author[0000-0002-7670-670X]{Malena Rice}
\affiliation{Department of Astronomy, Yale University, P.O. Box 208101, New Haven, CT 06520-8101, USA}
\email{malena.rice@yale.edu}

\author[0000-0002-1836-3120]{Avi Shporer}
\affiliation{MIT Kavli Institute for Astrophysics and Space Research, Massachusetts Institute of Technology, 70 Vassar Street, Cambridge, MA 02139, USA}
\email{shporer@space.mit.edu}

\author[0000-0002-8964-8377]{Samuel N. Quinn}
\affiliation{Center for Astrophysics | Harvard \& Smithsonian, 60 Garden Street, Cambridge, MA 02138, USA}
\email{squinn@cfa.harvard.edu}

\author[0009-0004-8891-4057]{Marcelo Tala Pinto}
\affiliation{Department of Astronomy, The Ohio State University, Columbus, OH 43210, USA}
\email{marcelo.tala@edu.uai.cl}

\author[0000-0003-2527-1475]{Shreyas Vissapragada}
\affiliation{Observatories of the Carnegie Institution for Science, 813 Santa Barbara Street, Pasadena, CA 91101, USA}
\email{svissapragada@carnegiescience.edu}

\author[0000-0001-7961-3907]{Samuel W. Yee}
\affiliation{Division of Astronomy and Astrophysics, University of California, Los Angeles, Los Angeles, CA, USA}
\email{syee@astro.ucla.edu}

\author[0000-0002-4891-3517]{George Zhou}
\affiliation{Centre for Astrophysics, University of Southern Queensland, Toowoomba, QLD 4350, Australia}
\email{george.zhou.astro@gmail.com}

\begin{abstract}

Cold Jupiters constrain theories of planet formation and dynamical evolution, but their long orbital periods complicate detection and characterization.
We combine long-baseline radial velocities (RVs) with Hipparcos intermediate astrometry and Gaia DR2/DR3 catalog measurements to investigate giant companions in three nearby systems.
Joint Markov chain Monte Carlo modeling with Gaussian-process RV covariance yields a period of ${20.26}_{-0.62}^{+0.65}$~yr and a mass of ${3.76}_{-0.30}^{+0.44}\,M_{\rm Jup}$ for HD~361~b, with a low eccentricity of ${0.07}_{-0.03}^{+0.04}$.
HD~196068~b, hosted by one component of a wide stellar binary, has a period of ${22.89}_{-0.82}^{+0.89}$~yr, an eccentricity of ${0.20}_{-0.04}^{+0.03}$, and a minimum mass of ${2.21}_{-0.15}^{+0.16}\,M_{\rm Jup}$; its inclination remains unconstrained.
For HD~203432, the adopted two-planet model yields periods of $17.35_{-0.43}^{+0.48}$~yr and $47_{-14}^{+27}$~yr, with corresponding masses of $0.66_{-0.18}^{+0.32}\,M_{\rm Jup}$ and $2.43_{-0.45}^{+0.48}\,M_{\rm Jup}$.
HD~203432~b and HD~203432~c have eccentricities of $0.63_{-0.12}^{+0.11}$ and $0.49\pm0.13$, respectively.
Astrometry yields prograde and retrograde orbital solutions.
HD~361~b provides a nearly circular counterpart to the eccentric HD~203432 pair, which includes a sub-Jovian companion beyond 6~au.
HD~196068~b establishes HD~196067--HD~196068 as a wide binary hosting giant planets around both stars, enabling a comparison of contrasting architectures within one binary.
Continued RV monitoring and time-resolved astrometry will refine the long-period orbits, companion masses, and three-dimensional architectures of these systems.

\end{abstract}

% UAT concepts are supplied during AAS submission.

\section{Introduction} 
% To date, more than 6000 exoplanets have been discovered and confirmed via RV, transit, direct imaging, astrometry and microlensing \citep{Akeson2013}.
% Among them, cold Jupiters play a crucial role in shaping the architecture and potential habitability of planetary systems, driving significant interest in understanding their formation, evolution, and dynamics.

The discovery of exoplanets over the past three decades has transformed our understanding of planetary systems and their formation. 
Exoplanets detected using a wide variety of observational techniques are compiled in the NASA Exoplanet Archive \citep{Christiansen2025}. 
Long-period giant planets, often referred to as ``cold Jupiters,'' remain comparatively underexplored because of the observational difficulty of detecting planets on wide orbits. 
Here we use ``cold Jupiters'' for giant planets with masses greater than $\sim 0.3 \ M_{\rm Jup}$ and semi-major axes exceeding $\sim 1$ AU, following the working definition of \citet{Zhu2018}. 
In the core accretion paradigm, they provide a critical test of whether solid cores can grow rapidly enough beyond the snow line to trigger runaway gas accretion before disk dispersal \citep{Pollack1996, Ida2004, Lambrechts2012}. 
Within the gravitational instability framework, cold Jupiters probe whether disks can undergo fragmentation under realistic cooling conditions \citep{Boss1997, Rafikov2005, Kratter2016}. 
Their occurrence rates, metallicity dependence, masses, and orbital architectures therefore provide important constraints for distinguishing between these two formation pathways \citep{Cui2026}. 
Moreover, cold Jupiters can affect the formation and dynamical evolution of inner low-mass planets \citep{Izidoro2015, Pu2021, Guo2023}, including their resonant architecture \citep{Guo2026}, and can influence volatile delivery to terrestrial-planet regions \citep{Raymond2004, Raymond2006, Ogihara2023, Kane2024}. 
Several statistical studies have reported an enhanced occurrence of outer giant planets in systems hosting inner small planets \citep{Zhu2018, Bryan2019, Rosenthal2022, VanZandt2025}, whereas \citet{Bonomo2023} found no evidence for such an excess in their sample.
Metallicity and stellar mass affect this comparison \citep{Zhu2024, Bryan2025}. A positive association has been reported for metal-rich Sun-like hosts \citep{Bryan2024}, but \citet{Bonomo2025} found only a possible association in the super-solar mass and metallicity bin, at less than $3\sigma$ confidence.
The relation between the two populations remains uncertain for other stellar types and metallicities because samples remain limited. 
Therefore, establishing a statistically robust sample of cold Jupiters is essential for understanding the broader context of planetary system evolution. 

Despite their importance, cold Jupiters remain difficult to detect. 
Transit surveys are intrinsically biased toward short-period systems because of the low geometric probability and infrequent transits associated with long orbital periods. 
Radial velocity (RV) observations, while highly sensitive to giant planets, require long observational baselines to fully characterize wide-orbit companions. 
Even when a periodic RV signal is detected, the inferred companion mass is subject to the well-known $M_{p} \sin i$ degeneracy, yielding only a minimum mass estimate. 
As a result, distinguishing between giant planets, brown dwarfs, and low-mass stellar companions can be challenging for systems viewed at low inclinations. 

Astrometry provides a powerful complementary technique that can resolve this ambiguity. 
% By measuring the reflex motion of the host star on the plane of the sky, astrometric observations directly constrain the orbital inclination and thus the true companion mass. 
For long-period massive planets, it is possible to combine Doppler measurements with high-precision space-based astrometry, revealing the full 3-D orbits and determining robust dynamical masses. 
In recent years, several groups have developed such techniques that combine Hipparcos and Gaia astrometry  \citep[e.g.,][]{Snellen2018, Brandt2018, Brandt2021, Feng2019, Kervella2019, Xuan2020, Venner2021, Sozzetti2023, VanZandt2024, Thompson2025, An2025, Wallace2026}. 
Joint analyses can constrain orbital inclination and companion mass in favourable cases, using proper-motion anomalies together with other orbital information.
Treating a catalog proper motion as an instantaneous measurement can bias orbital inference when the orbital period is comparable to the mission observing window \citep{Feng2023}.
Proper-motion anomaly fits can retain prograde--retrograde inclination alternatives \citep{Venner2021}. Combining multiple Gaia releases can help distinguish these alternatives in favourable cases \citep{Xiao2024}.

We investigate long-period giant companions in three systems by jointly modeling RV and astrometric observations, following \citet{Feng2023,Xiao2024}. 
Individual stellar epoch astrometry is expected with Gaia DR4\footnote{\url{https://www.cosmos.esa.int/web/gaia/dr4}}; here we instead simulate along-scan model predictions using the Gaia Observation Schedule Tool (GOST\footnote{\url{https://gaia.esac.esa.int/gost/}}).
GOST provides predicted transit times and scan angles for sampling the astrometric model\footnote{\url{https://www.cosmos.esa.int/web/gaia/tools}}. 
This method has been applied to several cold Jupiters, including $\epsilon$ Indi A b, $\epsilon$ Eridani b \citep{Feng2023, Feng2025}, HD 222237 b \citep{Xiao2024}, HD 68475 b, HD 100508 b, HD 48265 c, HD 114386 c \citep{Wu2026}, and HD 164604 c \citep{Xiao2026}.

In this paper, the RV data are collected from high-precision spectrographs including PFS \citep{Crane2006, Crane2008, Crane2010}, MIKE \citep{Bernstein2003}, HARPS \citep{Pepe2000} and UCLES \citep{Diego1990}. 
We combine those RVs with high-precision astrometry from Hipparcos and Gaia second and third releases (GDR2 and GDR3; \citealt{Gaia2018_DR2summary, Gaia2023_DR3summary}) to constrain the masses and orbital properties of cold Jupiters.

The paper is organized as follows.
Section~\ref{sec:data} describes the RV and astrometric observations, and Section~\ref{sec:stellar} presents the stellar characterization.
Section~\ref{sec:method} defines the joint models, and Section~\ref{sec:results} reports the orbital posteriors and model comparisons for the three systems.
Section~\ref{sec:discussion} discusses their physical interpretation, dynamical evolution, and observational prospects. Section~\ref{sec:conclusions} summarizes the main findings.

\section{Data} \label{sec:data}
We combine multi-instrument RVs with astrometry to constrain the orbits of long-period, massive companions. Instrument-specific RV offsets account for zero-point differences. 
Table~\ref{tab:RV_data_summary} summarizes the RV observations.

\begin{deluxetable*}{cccccc}
\tablewidth{0pt}
\tablecaption{Summary of RV data \label{tab:RV_data_summary}}
\tablehead{
\colhead{Instrument} & \colhead{Time span} & \colhead{Length} & \colhead{$N_{\rm{obs}}$} & \colhead{$\langle \sigma_{\rm{RV}} \rangle$} & \colhead{Data source} \\
\colhead{} & 
\colhead{(days)} & 
\colhead{} & 
\colhead{(m s$^{-1}$)} & 
\colhead{}}
\startdata
 & & \makecell*[c]{HD 361} & & \\
\hline
HARPSpre & Oct. 2003 -- Nov. 2013 & 3685 & 21 & 0.39 & ESO archive$^{a}$ \\
HARPSpost & Oct. 2021 -- Sept. 2023 & 704 & 62 & 0.52 & ESO archive \\
MIKE & Dec. 2002 -- Jul. 2009 & 2390 & 9 & 4.67 & This work \\
PFS & Jan. 2010 -- Sept. 2026 & 6086 & 77 & 1.52 & This work \\
\hline
 & & \makecell*[c]{HD 196068} & & \\
\hline
AAT & Nov. 1998 -- Sept. 2015 & 6169 & 36 & 2.15 & This work \\
PFS & Jun. 2012 -- Sept. 2026 & 5210 & 77 & 1.25 & This work \\
\hline
 & & \makecell*[c]{HD 203432} & & \\
\hline
HARPSpre & Oct. 2003 -- Sept. 2007 & 1405 & 74 & 0.90 & ESO archive \\
MIKE & Jun. 2003 -- Jul. 2009 & 2209 & 9 & 4.25 & This work \\
PFS & May 2010 -- Sept. 2026 & 5943 & 82 & 1.12 & This work \\
\enddata
\begin{flushleft}
$^{a}$ ESO archive website: \url{https://archive.eso.org/scienceportal/home}
\end{flushleft}
%\tablecomments{$^{a}$ ESO archive website: \url{https://archive.eso.org/scienceportal/home}}
%\tablenotetext{a}{\url{https://archive.eso.org/scienceportal/home}}
\end{deluxetable*}

\subsection{PFS}
The Carnegie Planet Finding Spectrograph (PFS) is an optical echelle spectrograph installed on the 6.5-meter Magellan II (Clay) Telescope at Las Campanas Observatory in Chile \citep{Crane2006, Crane2008, Crane2010}.
The current instrument specifications give a wavelength range of 391--734 nm and resolving powers from 38,000 to 190,000\footnote{\url{https://users.obs.carnegiescience.edu/crane/pfs/}}.
A molecular iodine absorption cell superimposes reference absorption features on the stellar spectra for point-spread-function deconvolution and wavelength calibration. The spectral data were reduced and RVs computed using a customized pipeline following \citet{Butler1996}.
PFS has been in scientific operation since 1 January 2010. 

Additionally, we extracted two common stellar activity indicators: the emission fluxes of the Ca~II H\&K lines and the H$\alpha$ line. Flux time series were derived following \citet{Perdelwitz2021_SHK}. The resulting time series are shown in Figure~\ref{fig:activity_diagnostics}.
% https://users.obs.carnegiescience.edu/crane/pfs/ 

\subsection{Other RV instruments}
The Anglo-Australian Planet Search (AAPS) began in 1998 on the 3.9-meter Anglo-Australian Telescope (AAT) in Australia \citep{Tinney2001}. 
Equipped with the University College London Echelle Spectrograph (UCLES; \citefull{Diego1990}) and combined with an iodine absorption cell for calibration \citep{Marcy1992}, it operated at spectral resolution $R \sim 45,000$, and maintained an RV precision of about $3 \ \rm{m \ s^{-1}}$ \citep{Tinney2005}.
The data reduction, including the recovery of the spectrometer point-spread-function, is described by \citet{Butler1996}.
% https://grokipedia.com/page/anglo_australian_planet_search

The High Accuracy Radial velocity Planet Searcher (HARPS; \citefull{Pepe2000}), which began operation in 2003, is a high-precision echelle planet-finding spectrograph mounted on ESO's 3.6-meter telescope at La Silla Observatory in Chile.
The HARPS spectrograph underwent a major fiber link upgrade at the end of May 2015 \citep{LoCurto2015}. 
We distinguish between the HARPSpre and HARPSpost data, treating them as separate datasets with individual zero-point offsets.
The HARPS-RVBANK archive provides public HARPS radial velocities reprocessed with the SERVAL pipeline \citep{Zechmeister2018,Trifonov2020}\footnote{\url{https://cdsarc.cds.unistra.fr/viz-bin/cat/J/A+A/636/A74}}.

The Magellan Inamori Kyocera Echelle (MIKE; \citefull{Bernstein2003}) is a double echelle spectrograph, installed on the 6.5-meter Magellan II (Clay) Telescope at Las Campanas Observatory in Chile.
It was installed on Clay in November 2002; its blue and red channels cover approximately 320--500 nm and 490--1000 nm, respectively \citep{Bernstein2015}.
Spectral resolution is 22,000 and 28,000 on the red and blue sides, respectively, with a 1 arcsec slit. 
The MIKE data reduction is described in detail by \citet{Bernstein2015}.
% https://www.lco.cl/?epkb_post_type_1=the-mike-magellan-inamori-kyocera-echelle-users-guide

\subsection{Hipparcos and Gaia}
Astrometry complements the line-of-sight information in RVs with constraints in the plane of the sky. 
One source of astrometric data is the Hipparcos mission, an astronomical satellite operating between 1989 and 1993. Equipped with a 0.29-meter telescope, it provided precise measurements of stellar positions and proper motions for approximately 118,200 stars \citep{ESA1997}.
Gaia was launched by ESA in 2013 \citep{Gaia2016} and completed its science observations in January 2025\footnote{\url{https://www.cosmos.esa.int/web/gaia/end-of-observations}}. 
Gaia employs a dual-telescope design with two rectangular primary mirrors, each measuring $1.45\,\mathrm{m}\times0.5\,\mathrm{m}$ \citep{Gaia2016}. Gaia DR2 provides five-parameter astrometry for about 1.3 billion sources, with precision that depends strongly on magnitude \citep{Gaia2018}.

In this work, we make use of astrometric data from both Hipparcos and Gaia catalogs, including stellar parallaxes ($\varpi$), positions (R.A. $\alpha$ and decl. $\delta$), and proper motions ($\mu_{\alpha*}\equiv\dot{\alpha}\cos\delta$ and $\mu_{\delta}$). 
In particular, we incorporate Hipparcos intermediate astrometry data (IAD; \citefull{vanLeeuwen2007}) and GDR2 and GDR3 catalog astrometry \citep{Gaia2018_DR2summary, Gaia2023_DR3summary}. 
We use GOST to predict transit epochs and scan-angle geometry, which are combined with our astrometric model to simulate along-scan predictions \citep{Feng2023,Xiao2024}. These predictions enter the catalog-level astrometric calculation.
The catalog astrometry and radial velocity observations are jointly modeled while accounting for zero-point offsets and jitter for each instrument. 
The long baseline ($\sim25 $yr) between Hipparcos and Gaia enables tighter constraints on the three-dimensional motion of host stars, helping to break the $M_{\mathrm{p}} \sin i$ degeneracy and improve orbital solutions for long-period, massive exoplanets.

\section{Stellar Characterization} \label{sec:stellar}
HD~361 and HD~196068 are classified as G1V, and HD~203432 as G8IV \citep{Gray2006}. All three targets have long-term RV monitoring and precision spectroscopy.
As part of the PFS data reduction pipeline, we constructed an iodine-free stellar template spectrum from high-SNR observations of each star, using bracketing B-star observations through the iodine cell for precise wavelength calibration. 
We first determined the effective temperatures ($T_{\rm eff}$) and metallicities ([Fe/H]) from the PFS template spectra using \texttt{SpecMatch-Emp} \citep{Yee2017}. 
\texttt{SpecMatch-Emp} compares an input spectrum against a library of Keck/HIRES spectra, identifies the five best-matching library stars, and interpolates their stellar parameters to estimate parameters of the target star. We supplied the 500--650 nm region of the PFS spectra to \texttt{SpecMatch-Emp}.

The stars have effective temperatures between $5400 \ \rm{K}$ and $6000 \ \rm{K}$ and metallicities [Fe/H] ranging from $-0.20$ to $0.35$. 
% For those three G-type stars, we adopt their effective temperature, $\log g$ and [Fe/H] from \cite{Soubiran2022}, and derive their stellar masses, radii and ages via isochrone fitting using \texttt{isochrones} \citep{Morton2015}. 
% For those three target stars, we determine their effective temperature, $\log g$ and [Fe/H] from PFS spectra using \texttt{specmatch-emp} \citep{Yee2017}, and then infer their stellar masses and ages through isochrone fitting with \texttt{SEDEX\_ISO}\footnote[3]{\url{https://github.com/Jiayue-Zhang99/SEDEX_ISO/}}. 

%\subsection{SED and isochrone fitting}
We constrained the broadband spectral energy distribution (SED) and host-star evolutionary properties using our pipeline \texttt{SEDEX\_ISO}\footnote{\url{https://github.com/Jiayue-Zhang99/SEDEX_ISO.git}}, which extends SEDEX \citep{Yu2023_SEDEX} by coupling SED fitting to isochrone-based inference. Multiband photometry was assembled automatically from archival catalogs, including Gaia DR3 \citep{Riello2021_GaiaPhot}, Tycho-2 \citep{hog2000_Tycho}, Hipparcos \citep{vanLeeuwen2007}, 2MASS \citep{Cutri2003_2MASS,Skrutskie2006_2MASS,Skrutskie2003_2MASS_IPAC}, ALLWISE \citep{Cutri2014_ALLWISE,Wright2010_WISE,Wright2019_AllWISE_IPAC}, and SkyMapper \citep{Onken2019_SkyMapper}, as available for each target.

For the SED analysis, we fitted the broadband photometry with high-resolution synthetic spectra from the MARCS library \citep{Gustafsson2008}. We adopted the spectroscopic effective temperature and metallicity from \texttt{SpecMatch-Emp}, together with Gaia DR3 GSP-Phot surface-gravity priors \citep{Andrae2023}, as external constraints, and incorporated the Gaia-based distance in the inference. The fit yields posterior estimates of the bolometric flux scale, angular radius, broadband extinction, and atmospheric parameters, including $T_{\rm eff}$ and an SED-based surface gravity ($\log g_{\rm SED}$).

These SED-derived atmospheric parameters were then passed to an isochrone analysis with the \texttt{isochrones} package \citep{Morton2015} and MIST model grids \citep{Dotter2016_MIST1,Choi2016_MIST2}, which provides posterior estimates of the stellar mass, radius, mean density, luminosity, age, and an isochrone-based surface gravity ($\log g_{\rm MIST}$). Because broadband SED fitting only weakly constrains $\log g$, whereas $\log g_{\rm MIST}$ is obtained jointly with the mass and radius from the same evolutionary-model posterior, we adopt $\log g_{\rm MIST}$ as the preferred surface gravity. In Table~\ref{tab:host_stars} we report the SED-inferred $T_{\rm eff}$, the spectroscopic [Fe/H] from \texttt{SpecMatch-Emp}, and the isochrone-inferred mass, age, and $\log g_{\rm MIST}$.

The SED fits for HD~361 and HD~203432 are of good quality (reduced $\chi^2\simeq0.8$ and $0.2$, respectively). By contrast, HD~196068 returns a substantially larger reduced $\chi^2$ ($\simeq7$) and a formally negative extinction, indicating tension among the available broadband photometry. This photometric tension limits the reliability of the SED scale and extinction for HD~196068 and introduces systematic uncertainty in its isochrone-derived mass and age.%; for this star the \texttt{SpecMatch-Emp} $T_{\rm eff}$ and [Fe/H] remain the more robust spectroscopic anchors.

The weighted PASTEL means compiled by \citet{Soubiran2022} provide a spectroscopic comparison. For HD~361, HD~196068, and HD~203432, respectively, they give $T_{\rm eff}=5900\pm20$, $5909\pm53$, and $5629\pm20$~K, with $[\mathrm{Fe/H}]=-0.12\pm0.02$, $0.26\pm0.07$, and $0.29\pm0.03$~dex. Our SED-based temperatures are approximately $130$--$140$~K lower, while the metallicities retain the same contrast between the mildly metal-poor HD~361 and the metal-rich HD~196068 and HD~203432. The PASTEL surface gravity of HD~203432, $\log g=4.36\pm0.05$, agrees with our isochrone estimate of $4.31\pm0.11$.

Table~\ref{tab:host_stars} summarizes the stellar parameters.

\begin{deluxetable*}{ccccc}
\tablewidth{0pt}
\tablecaption{Stellar parameters of host stars investigated in this work \label{tab:host_stars}}
\tablehead{
Name & HD 361 & HD 196068 & HD 203432 & Source}
\startdata
HIP ID & 669 & 102128 & 105606 & 1 \\
Gaia DR3 ID & 2417456431710276096 & 6369544118966055296 & 6582020205549054720 & 2 \\
Spectral Type & G1V & G1V & G8IV & 3 \\
$\alpha_{\rm J2016.0}$ & $00^{\rm h}08^{\rm m}16.35^{\rm s}$ & $20^{\rm h}41^{\rm m}46.21^{\rm s}$ & $21^{\rm h}23^{\rm m}22.46^{\rm s}$ & 2 \\
$\delta_{\rm J2016.0}$ & $-14^\circ 49' 28.35''$ & $-75^\circ 20' 49.79''$ & $-41^\circ 33' 11.36''$ & 2 \\
$\varpi$ (mas) & $36.617 \pm 0.024$ & $25.038 \pm 0.017$ & $28.482 \pm 0.030$ & 2 \\
$\mu_{\alpha*}$ (mas/yr) & $-14.371 \pm 0.029$ & $163.531 \pm 0.016$ & $-73.088 \pm 0.024$ & 2 \\
$\mu_{\delta}$ (mas/yr) & $-11.240 \pm 0.020$ & $-171.346 \pm 0.018$ & $-144.509 \pm 0.020$ & 2 \\
$u_{\rm SM}$ (mag) & \nodata & \nodata & $9.604 \pm 0.061$ & 4 \\
$v_{\rm SM}$ (mag) & \nodata & \nodata & $9.206 \pm 0.061$ & 4 \\
$B_{T}$ (mag) & $7.780 \pm 0.025$ & $7.790 \pm 0.025$ & $8.553 \pm 0.026$ & 5 \\
$V_{T}$ (mag) & $7.103 \pm 0.022$ & $7.070 \pm 0.022$ & $7.686 \pm 0.023$ & 5 \\
$H_{p}$ (mag) & $7.170 \pm 0.020$ & $7.300 \pm 0.044$ & $7.753 \pm 0.020$ & 6 \\
$G$ (mag) & $6.8956 \pm 0.0002$ & $7.0112 \pm 0.0002$ & $7.4392 \pm 0.0002$ & 2 \\
$G_{\rm{BP}}$ (mag) & $7.1977 \pm 0.0007$ & $7.3170 \pm 0.0006$ & $7.7968 \pm 0.0004$ & 2 \\
$G_{\rm{RP}}$ (mag) & $6.4209 \pm 0.0006$ & $6.5364 \pm 0.0005$ & $6.9179 \pm 0.0004$ & 2 \\
$J$ (mag) & $5.947 \pm 0.034$ & $6.061 \pm 0.034$ & $6.374 \pm 0.028$ & 7 \\
$H$ (mag) & $5.628 \pm 0.047$ & $5.796 \pm 0.039$ & $6.063 \pm 0.043$ & 7 \\
$K_{S}$ (mag) & $5.556 \pm 0.039$ & $5.713 \pm 0.027$ & $5.935 \pm 0.034$ & 7 \\
$W1$ (mag) & $5.508 \pm 0.164$ & \nodata & \nodata & 8 \\
$W3$ (mag) & $5.626 \pm 0.035$ & \nodata & \nodata & 8 \\
$T_{\rm{eff}}$ (K) & $5770 \pm 108$ & $5771 \pm 103$ & $5493 \pm 95$ & 9 \\
$\log g \ (\log_{10}(\mathrm{cm\,s^{-2}}))$ & $4.36 \pm 0.13$ & $4.25 \pm 0.07$ & $4.31 \pm 0.11$ & 9 \\
$\rm{[Fe/H] \ (dex)}$ & $-0.18 \pm 0.09$ & $0.25 \pm 0.09$ & $0.33 \pm 0.09$ & 9 \\
$M_{*} (M_{\odot})$ & $0.96 \pm 0.03$ & $1.16 \pm 0.03$ & $1.04 \pm 0.04$ & 9 \\
Age (Gyr) & $3.26 \pm 1.60$ & $4.62 \pm 1.06$ & $5.65 \pm 2.32$ & 9 \\
\enddata
\begin{flushleft}
\tablerefs{
[1] \cite{ESA1997},
[2] \cite{Gaia2023_DR3summary},
[3] \cite{Gray2006},
[4] \cite{Onken2019_SkyMapper},
[5] \cite{hog2000_Tycho},
[6] \cite{vanLeeuwen2007},
[7] \cite{Cutri2003_2MASS},
[8] \cite{Cutri2014_ALLWISE},
[9] this work.}

\par Photometric uncertainties are the values adopted in this work and are not all identical to the formal catalog errors.
\end{flushleft}
\end{deluxetable*}

\section{Method} \label{sec:method}
We adopt a right-handed coordinate system defined by the orthonormal triad $[\hat{x}, \hat{y}, \hat{z}]$, where $\hat{x}$ points in the direction of increasing R.A., $\hat{y}$ points in the direction of increasing decl., and $\hat{z}$ points away from the observer (convention $\mathrm{III}$ of \citealt{PEXO2019}). 
This convention is used in the astrometry models of Hipparcos and Gaia \citep{ESA1997, Lindegren2012}. 

We jointly model the RVs, Hipparcos IAD, and Gaia DR2/DR3 catalogs using the forward-modeling framework of \citet{Feng2023} and \citet{Xiao2024}. For each Gaia release, the model is sampled at the GOST transit epochs and scan angles and reduced to a synthetic five-parameter solution over that release's observing window.
The primary fitted parameters in our model include seven orbital elements (orbital period $P$, RV semi-amplitude $K$, two parameters relevant to eccentricity and argument of periastron $\sqrt{e}\cos \omega$ and $\sqrt{e}\sin \omega$, orbital inclination $i$, longitude of ascending node $\Omega$, mean anomaly $M_{0}$ at epoch J2017.5 (JD2,457,936.875)), five astrometric offsets ($\Delta \alpha_*, \Delta \delta, \Delta \varpi, \Delta \mu_{\alpha*}$ and $\Delta \mu_{\delta}$) of barycenter relative to GDR3, RV offsets $\gamma$ and jitter terms $\sigma_{\rm jit}$ for each RV instrument, as well as an astrometric jitter $J_{\rm hip}$ for Hipparcos IAD, and an error inflation factor $S_{\rm gaia}$ for Gaia astrometry. 
The semi-major axis $a$ of the planet relative to the host, the mass of planet $m_{p}$, the planet-to-host mass ratio $q$, and the epoch of periastron passage $T_{p}$ can be derived from the above orbital elements.

For the RV model, the likelihood function is expressed as
\begin{equation}
\begin{split}
  \mathcal{L}_{\rm RV}
  &=\prod\limits_{k=1}^{N_{\rm inst}}\prod\limits_{j=1}^{N_{{\rm RV},k}}
  \frac{1}{\sqrt{2\pi(\sigma_{j,k}^2+\sigma_{{\rm jit},k}^2)}} \times\\
  &\quad \exp\!\left( -\frac{(v_{j,k}-\hat{v}_{j,k}-\gamma_k)^2}
    {2(\sigma_{j,k}^2+\sigma_{{\rm jit},k}^2)} \right)~,
\end{split}
\end{equation}
where $N_{{\rm RV},k}$ is the number of measurements from instrument $k$, $N_{\rm RV}=\sum_k N_{{\rm RV},k}$, and $N_{\rm inst}$ is the number of instruments. 
$\gamma_k$ denotes the RV offset for the $k$-th instrument, and $\sigma_{{\rm jit},k}$ represents the jitter term to account for unmodeled stellar or instrumental noise. 
$v_{j,k}$ and $\hat{v}_{j,k}$ are the observed and predicted RVs, respectively.

Stellar activity can produce RV variations that are correlated in time, motivating the use of a Gaussian process (GP) to describe this variability alongside the Keplerian signals \citep{Rajpaul2015}. For HD~361 and HD~203432, the GP models time-correlated RV variations within each instrument, while independent instrumental jitter describes additional white-noise scatter. HD~196068 uses the independent white-noise RV model. Here we use a phenomenological Mat\'ern-$3/2$ covariance \citep[Equation~4.17]{Rasmussen2006},
\begin{equation}
\label{eq:gp_covariance}
C_{ij}=A_{\rm GP}^{2}(1+x_{ij})e^{-x_{ij}}
 +(\sigma_i^2+\sigma_{{\rm jit},k}^2)\delta_{ij},
\end{equation}
where $x_{ij}=\sqrt{3}|t_i-t_j|/\tau_{\rm GP}$ for measurements from instrument $k$. Here $A_{\rm GP}$ is the correlated RV amplitude and $\tau_{\rm GP}$ is its decay timescale. These hyperparameters are shared between instruments, whose covariance blocks are independent; each instrument retains its own white-noise jitter, and its RV offset is marginalized in the likelihood. The kernel is non-periodic.

In the RV figures, instrumental offsets are subtracted from the measurements. For HD~361 and HD~203432, the GP conditional posterior mean at the observing epochs is also subtracted. The black curves show posterior-mean Keplerian predictions, and the shaded bands enclose their pointwise 16th--84th percentile intervals. Residuals subtract the complete orbital, GP, and instrument-offset prediction. The plotted error bars combine measurement uncertainties and the best-fit instrumental jitter in quadrature.

The likelihood for Hipparcos IAD is
\begin{equation}
  \mathcal{L}_{\rm hip}=\prod\limits_{j=1}^{N_{\rm IAD}}\frac{1}{\sqrt{2\pi(\sigma_{j}^2+J_{\rm hip}^2)}}
  \times 
  \exp\!\left( -\frac{(\xi_j-\hat{\xi}_j)^2}{2(\sigma_{j}^2+J_{\rm hip}^2)} \right)~,
\end{equation}
where $N_{\rm IAD}$ is the number of epochs of Hipparcos IAD, $\sigma_j$ is the individual measurement uncertainty, and $J_{\rm hip}$ is the jitter term. 
The observed and predicted one-dimensional abscissae are denoted by $\xi_j$ and $\hat{\xi}_j$, respectively. 
We obtain $\hat{\xi}_j$ by projecting the combined barycentric, stellar reflex, and parallactic motions onto the Hipparcos scan direction. 
The model for $\hat{\xi}_j$ also includes the offset between Hipparcos astrometry and the solution propagated from the GDR3 epoch to the Hipparcos epoch.

The likelihood function for GDR2 and GDR3 can be written as
\begin{equation}
\begin{split}
  \mathcal{L}_{\rm gaia}
  &=\prod_{k=1}^{N_{\rm DR}}\frac{\exp[-\tfrac{1}{2}\mathbf{r}_k^{T}\mathbf{C}_k^{-1}\mathbf{r}_k]}{\sqrt{(2\pi)^5|\mathbf{C}_k|}},\\
  \mathbf{C}_k&\equiv S_{\rm gaia}^{2}\boldsymbol{\Sigma}_{k,\rm cat}.
\end{split}
\end{equation}
where $\mathbf{r}_k=\Delta\vec{\iota}_k-\Delta\hat{\vec{\iota}}_k$ and $N_{\rm DR}=2$ for the Gaia DR2 and DR3 catalogs.
For the Gaia likelihood, catalog positions are expressed relative to the fixed GDR3 position origin, while parallaxes and proper motions retain their absolute catalog values. We define
\begin{equation}
\begin{split}
\Delta\vec{\iota}_k\equiv
&\,((\alpha_k -\alpha_{\rm DR3})\cos\delta_k,\,\delta_k-\delta_{\rm DR3},\,\\
&\varpi_k-\varpi_{\rm DR3},\,\mu_{\alpha k}-\mu_{\alpha \rm DR3},\,\mu_{\delta k}-\mu_{\delta \rm DR3}),
\end{split}
\end{equation}
where $\Delta\hat{\vec{\iota}}_k$ is the synthetic five-parameter astrometric vector obtained by fitting the standard five-parameter model to the simulated abscissae, using the same position origin. Positions and parallaxes are expressed in mas, and proper motions in mas\,yr$^{-1}$. Here $\boldsymbol{\Sigma}_{k,\rm cat}$ is the full five-parameter catalog covariance matrix, and $S_{\rm gaia}$ is the shared uncertainty-inflation factor. The likelihood treats GDR2 and GDR3 as independent after frame calibration, with a shared uncertainty-inflation factor. 
We implement the frame rotation calibration between GDR2 and GDR3 catalog data recommended by \citet{Feng2024}\footnote{\url{https://github.com/ruiyicheng/Download_HIP_Gaia_GOST}}.

We adopt an isotropic prior on orbital inclination, equivalent to a uniform prior in $\cos i$. Stellar-mass uncertainty is propagated in the derived masses and semimajor axes at fixed orbital posterior, as described in Section~\ref{sec:results}. 
The total likelihood is $\mathcal{L}=\mathcal{L}_{\rm RV}\cdot\mathcal{L}_{\rm hip}\cdot\mathcal{L}_{\rm gaia}$. We sample the posterior using the parallel-tempering Markov chain Monte Carlo (MCMC) sampler \texttt{ptemcee} \citep{Vousden2016}.
Best-fit solutions denote the sampled parameter vectors with the highest posterior density (sampled maximum a posteriori, MAP, solutions); marginal distributions are summarized by their medians and equal-tail intervals. 
We begin with an RV-only fit to obtain rough estimates of $P$, $K$, $e$, $\omega$ and $M_0$. 
These estimates initialize the subsequent joint fit, while the initial values of $i$, $\Omega$, $\Delta \alpha_*$, $\Delta \delta$, $\Delta \mu_{\alpha*}$ and $\Delta \mu_\delta$ are drawn from uniform initialization distributions to explore possible bimodal solutions. 

Model comparison uses the Bayesian information criterion (BIC; \citealt{Schwarz1978}),
\begin{equation}
\label{eq:BIC}
\mathrm{BIC}=k\ln N-2\ln\hat{\mathcal{L}},
\end{equation}
where $k$ is the number of estimated parameters and $\hat{\mathcal{L}}$ is the maximized, normalized joint likelihood. For this calculation, the instrumental RV offsets are analytically profiled, and the GP enters through its covariance matrix. The parameter count includes the orbital elements, RV offsets and jitters, GP hyperparameters, barycentric astrometric corrections, and astrometric noise parameters. The observation count is $N=N_{\rm RV}+N_{\rm IAD}+10$, including the two five-dimensional Gaia catalog vectors. We define $\Delta\mathrm{BIC}=\mathrm{BIC}-\mathrm{BIC}_{\min}$ within each system; smaller values favor a model.

\section{Results} \label{sec:results}
We characterize the long-period companions in three systems, HD~361, HD~196068, and HD~203432, using the PFS radial velocities together with archival RV and Hipparcos--Gaia measurements. The labels 1p and 2p denote models with one and two Keplerian signals, respectively. We adopt a GP+1p model for HD~361, a white-noise+1p model for HD~196068, and a GP+2p model for HD~203432. Together, these models describe four companions. The PFS RV, activity, and residual periodograms are shown in Figure~\ref{fig:activity_diagnostics}. The strongest residual GLS peaks occur near $1.20$, $1.60$, and $1.32$~d for HD~361, HD~196068, and HD~203432, respectively, and lie below the plotted nominal 1\% white-noise false-alarm reference level.

Table~\ref{tab:planetary_paras} lists posterior medians and 16th--84th percentile equal-tail intervals. The posterior distributions include alternative period and orientation solutions. Planet masses and semimajor axes are recalculated sample by sample using the stellar masses in Table~\ref{tab:host_stars}, with independent Gaussian propagation of their quoted $1\sigma$ uncertainties. We denote the edge-on solution of the mass function by $M_{\min}$, conventionally written $M_p\sin i$ in the planetary-mass limit. The orientation entries in Table~\ref{tab:planetary_paras} summarize each companion's inclination hemispheres separately.

\begin{deluxetable*}{lccccc}[!ht]
\tabletypesize{\footnotesize}
\tablewidth{0pt}
\tablecaption{Posterior medians and 16th--84th percentile equal-tail intervals. HD~361 uses GP+1p, HD~196068 uses white noise+1p, and HD~203432 uses GP+2p. \label{tab:planetary_paras}}
\tablehead{Parameter & HD~361 b & HD~196068 b & HD~203432~b & HD~203432~c & Prior}
\startdata
Fitted parameters \\
\hline
Orbital Period $P$ (d) & ${7400}_{-226}^{+237}$ & ${8360}_{-299}^{+326}$ & ${6335}_{-156}^{+175}$ & ${17049}_{-5122}^{+9831}$ & Log - $\mathcal{U}(-10,16)$ \\
RV semi-amplitude $K$ (m/s) & ${36.5}_{-1.3}^{+1.3}$ & ${20.5}_{-1.3}^{+1.4}$ & ${7.6}_{-1.8}^{+2.2}$ & ${17.9}_{-3.5}^{+4.1}$ & Log - $\mathcal{U}(-10,16)$ \\
$\sqrt{e} \sin \omega$ & ${0.21}_{-0.11}^{+0.08}$ & ${-0.38}_{-0.06}^{+0.07}$ & ${0.11}_{-0.28}^{+0.24}$ & ${-0.40}_{-0.16}^{+0.20}$ & $\mathcal{U}(-1,1)$ \\
$\sqrt{e} \cos \omega$ & ${0.12}_{-0.11}^{+0.10}$ & ${0.23}_{-0.06}^{+0.06}$ & ${0.75}_{-0.12}^{+0.07}$ & ${0.55}_{-0.14}^{+0.13}$ & $\mathcal{U}(-1,1)$ \\
Inclination $i_j$ (deg), $i_j<90^\circ$ & ${63}_{-10}^{+15}$ & --- & ${62}_{-23}^{+19}$ & ${60}_{-16}^{+18}$ & $\cos i_j\sim\mathcal{U}(-1,1)$ \\
\hspace{1em}$i_j>90^\circ$ & ${106}_{-11}^{+13}$ & --- & ${120}_{-21}^{+24}$ & ${109}_{-14}^{+18}$ & --- \\
Node $\Omega_j$ (deg), $i_j<90^\circ$ & ${28.6}_{-8.1}^{+9.2}$ & --- & ${140}_{-86}^{+97}$ & ${34}_{-29}^{+313}$ & $\mathcal{U}(0,2\pi)$ \\
\hspace{1em}$i_j>90^\circ$ & ${33}_{-18}^{+326}$ & --- & ${126}_{-85}^{+117}$ & ${311}_{-20}^{+19}$ & --- \\
Mean anomaly at J2017.5 $M_0$ (deg) & ${45}_{-26}^{+31}$ & ${70.5}_{-8.9}^{+8.4}$ & ${213.8}_{-11}^{+7.9}$ & ${49}_{-19}^{+23}$ & $\mathcal{U}(0,2\pi)$ \\
\hline
Derived parameters \\
\hline
Orbital Period $P$ (yr) & ${20.26}_{-0.62}^{+0.65}$ & ${22.89}_{-0.82}^{+0.89}$ & ${17.35}_{-0.43}^{+0.48}$ & ${47}_{-14}^{+27}$ & --- \\
Semimajor axis $a$ (au) & ${7.35}_{-0.17}^{+0.17}$ & ${8.47}_{-0.22}^{+0.23}$ & ${6.78}_{-0.15}^{+0.15}$ & ${13.1}_{-2.8}^{+4.7}$ & --- \\
$M_p\sin i$ ($M_{\rm Jup}$) & ${3.41}_{-0.14}^{+0.15}$ & ${2.21}_{-0.15}^{+0.16}$ & ${0.55}_{-0.13}^{+0.15}$ & ${2.03}_{-0.46}^{+0.55}$ & --- \\
$M_p$ ($M_{\rm Jup}$) & ${3.76}_{-0.30}^{+0.44}$ & --- & ${0.66}_{-0.18}^{+0.32}$ & ${2.43}_{-0.45}^{+0.48}$ & --- \\
Eccentricity $e$ & ${0.07}_{-0.03}^{+0.04}$ & ${0.20}_{-0.04}^{+0.03}$ & ${0.63}_{-0.12}^{+0.11}$ & ${0.49}_{-0.13}^{+0.13}$ & --- \\
Argument of periastron $\omega$ (deg) & ${63}_{-30}^{+27}$ & ${300.7}_{-8.9}^{+10}$ & ${26}_{-18}^{+323}$ & ${323}_{-18}^{+18}$ & --- \\
Periastron time $T_p$ (JD$-2450000$) & ${7010}_{-610}^{+525}$ & ${6295}_{-176}^{+202}$ & ${4176}_{-124}^{+133}$ & ${5586}_{-286}^{+320}$ & --- \\
\enddata
\tablecomments{The equal-weight posterior samples include an additional period mode of HD~203432~b near $4200$~d, described below. For each companion $j$, the $(i_j,\Omega_j)$ entries are conditional on its own inclination hemisphere. For HD~203432, the conditions apply independently to $i_b$ and $i_c$ in their respective columns. Angular intervals are percentiles in linear degree coordinates, including intervals spanning the angular boundary. 
Planet masses and semimajor axes incorporate the stellar masses in Table~\ref{tab:host_stars} through independent Gaussian $1\sigma$ propagation at fixed orbital posterior.
Minimum masses solve the edge-on mass function (conventionally $M_p\sin i$). HD~196068 has an unconstrained orientation; its semimajor axis is evaluated at minimum mass. HD~203432~b and HD~203432~c denote the inner and outer companions, respectively. Log-$\mathcal{U}$ denotes a logarithmic-uniform prior; a prior uniform in $\cos i$ gives an isotropic distribution of orbital orientations.}
\end{deluxetable*}

\begin{deluxetable*}{lccc}[!ht]
\tabletypesize{\footnotesize}
\tablewidth{0pt}
\tablecaption{Correlated and white-noise parameters of the adopted joint models.\label{tab:GP_hyperparameters}}
\tablehead{Parameter & HD~361 & HD~196068 & HD~203432}
\startdata
GP amplitude $A_{\rm GP}$ (m s$^{-1}$) & ${5.82}_{-0.43}^{+0.48}$ & --- & ${3.21}_{-0.30}^{+0.37}$ \\
GP decay time $\tau_{\rm GP}$ (d) & ${2.52}_{-0.51}^{+0.68}$ & --- & ${1.88}_{-0.64}^{+0.65}$ \\
$\sigma_{\rm jit,HARPSpost}$ (m s$^{-1}$) & ${0.88}_{-0.63}^{+0.87}$ & --- & --- \\
$\sigma_{\rm jit,HARPSpre}$ (m s$^{-1}$) & ${0.36}_{-0.26}^{+0.54}$ & --- & ${0.90}_{-0.23}^{+0.27}$ \\
$\sigma_{\rm jit,MIKE}$ (m s$^{-1}$) & ${20.3}_{-5.4}^{+8.6}$ & --- & ${11.7}_{-3.3}^{+4.9}$ \\
$\sigma_{\rm jit,AAT}$ (m s$^{-1}$) & --- & ${6.23}_{-0.83}^{+1.03}$ & --- \\
$\sigma_{\rm jit,PFS}$ (m s$^{-1}$) & ${0.58}_{-0.39}^{+0.44}$ & ${1.49}_{-0.21}^{+0.22}$ & ${0.43}_{-0.29}^{+0.35}$ \\
$J_{\rm hip}$ (mas) & ${2.16}_{-0.26}^{+0.28}$ & ${89.1}_{-5.0}^{+5.5}$ & ${0.69}_{-0.48}^{+0.58}$ \\
$S_{\rm gaia}$ & ${1.19}_{-0.08}^{+0.08}$ & ${1.10}_{-0.06}^{+0.08}$ & ${1.10}_{-0.07}^{+0.08}$ \\
\enddata
\tablecomments{Medians and 16th--84th percentile equal-tail intervals. The Mat\'ern-$3/2$ covariance in Equation~\ref{eq:gp_covariance}, adopted for HD~361 and HD~203432, has shared GP hyperparameters and independent instrument covariance blocks. HD~196068 uses white noise with instrumental jitters. $J_{\rm hip}$ is the additive Hipparcos jitter; $S_{\rm gaia}$ is the dimensionless Gaia uncertainty-inflation factor.}
\end{deluxetable*}

\subsection{HD~361}
HD~361 is a G1V star \citep{Gray2006} at a distance of approximately $27.3$~pc, based on the parallax in Table~\ref{tab:host_stars}. We adopt our stellar mass of $0.96\pm0.03\,M_{\odot}$ and metallicity of $[\mathrm{Fe/H}]=-0.18\pm0.09$ from the same table.

\begin{figure*}[!ht]
\centering
\includegraphics[width=0.96\textwidth,height=0.72\textheight,keepaspectratio]{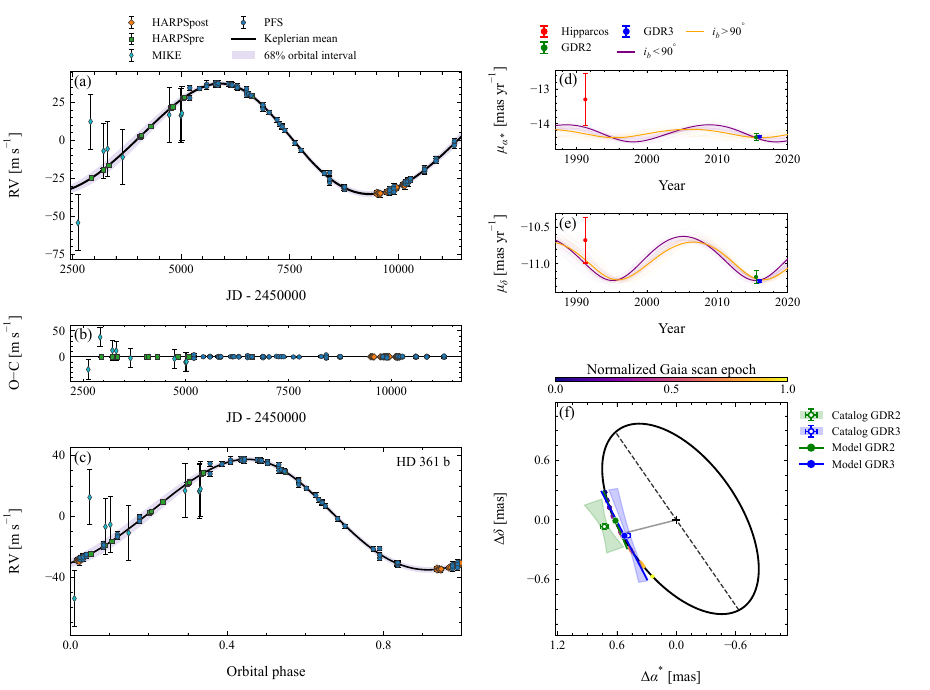}
\caption{Radial velocities and astrometric predictions for HD~361 under the adopted GP+1p model. (a) RV time series. (b) Full-model residuals. (c) Phase-folded velocities of HD~361~b. Instrument offsets and the chronological GP conditional posterior mean are subtracted from the RVs. Black curves show posterior-mean Keplerian signals; pale-purple shading encloses their pointwise 16th--84th percentile orbital intervals. RV error bars combine measurement errors and best-fit instrument jitter in quadrature. Residuals refer to the complete Keplerian+GP model. (d,e) Absolute proper-motion predictions in right ascension and declination. Purple and orange curves show the best-fit solutions in the two inclination hemispheres; thin curves show 100 posterior predictions. Points denote Hipparcos and Gaia DR2/DR3 proper motions with formal errors. (f) Stellar reflex orbit and Gaia catalog projections. Open circles with error bars show catalog positions relative to the best-fit barycentric motion; filled circles and short lines show model catalog projections. Shaded regions show the positional uncertainties obtained by propagating the formal catalogue position and proper-motion uncertainties across the Gaia observing windows.
% Shading propagates formal catalog position and proper-motion errors over the observing windows. 
The colour scale marks normalized Gaia scan epoch.}
\label{fig:HD361_RV}
\label{fig:HD361_Ast_fit}
\end{figure*}

Figure~\ref{fig:HD361_RV} shows the RV and astrometric fits. The GP+1p model yields an orbital period of ${20.26}_{-0.62}^{+0.65}$~yr, an RV semi-amplitude of ${36.5}_{-1.3}^{+1.3}$~m\,s$^{-1}$, and a low eccentricity of ${0.07}_{-0.03}^{+0.04}$. The corresponding semimajor axis is ${7.35}_{-0.17}^{+0.17}$~au and the minimum mass is $M_{\min}={3.41}_{-0.14}^{+0.15}$~$M_{\rm Jup}$.

The GP amplitude and decay time are reported with the instrumental jitters in Table~\ref{tab:GP_hyperparameters}. GP+1p has a lower BIC than white noise+1p by $97.7$ (Table~\ref{tab:RV_models}).

Both inclination hemispheres remain populated. Their conditional inclination summaries are ${63}_{-10}^{+15}$~deg for $i<90^\circ$ and ${106}_{-11}^{+13}$~deg for $i>90^\circ$; the associated nodal distributions are reported separately in Table~\ref{tab:planetary_paras}. The nodal posterior spans the $0^\circ/360^\circ$ boundary. The companion mass marginalized over both inclination hemispheres is ${3.76}_{-0.30}^{+0.44}$~$M_{\rm Jup}$ after stellar-mass propagation. These results indicate a long-period giant companion on a low-eccentricity orbit, with prograde and retrograde orientation solutions.

\subsection{HD~196068}
HD~196068 is the fainter component of the visual binary HD~196067--HD~196068. The parallax in Table~\ref{tab:host_stars} corresponds to a distance of approximately $39.94$~pc. We adopt $T_{\rm eff}=5771\pm103$~K and $M_\star=1.16\pm0.03\,M_{\odot}$ from our stellar characterization.

\begin{figure*}[!ht]
\centering
\includegraphics[width=0.98\textwidth,height=0.60\textheight,keepaspectratio]{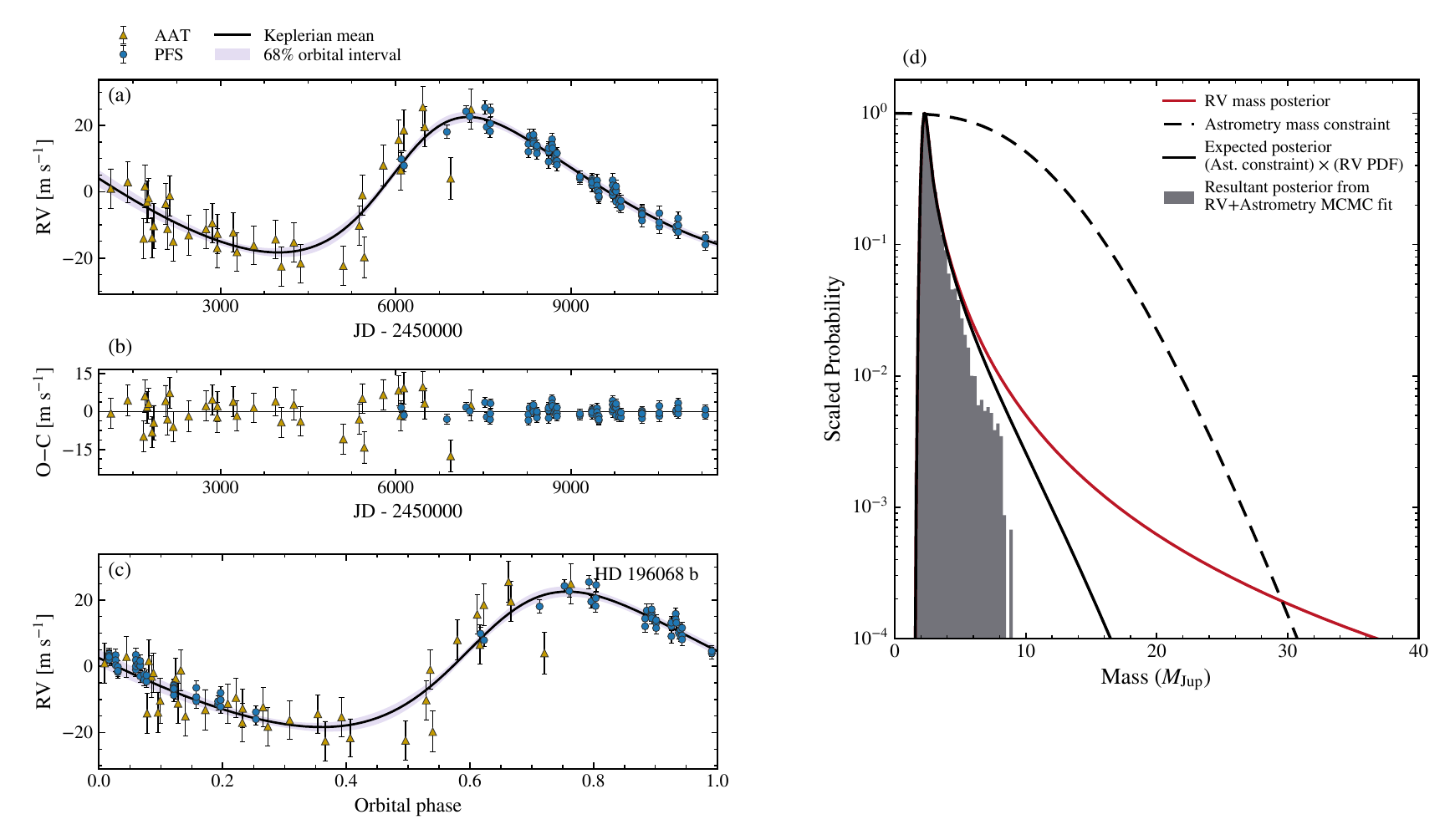}
\caption{Radial velocities and a supplementary mass diagnostic for HD~196068~b. Panels (a--c) show the adopted white-noise+1p fit: (a) RV time series, (b) model residuals, and (c) phase-folded velocities. Instrumental offsets are subtracted from the RVs. Black curves show posterior-mean Keplerian signals; pale-purple shading encloses their pointwise 16th--84th percentile orbital intervals. RV error bars combine measurement errors and best-fit instrumental jitter in quadrature. Panel (d) adapts the four-component presentation of Figure~6 in \citet{Errico2022}, using Hipparcos IAD and Gaia DR2/DR3 astrometry. Red: a separate RV-only posterior combined with isotropic orbital orientations. Dashed black: the independent astrometric mass constraint averaged over RV-informed trial orbits. Solid black: the normalized product of the red distribution and the astrometric constraint. Gray: the retained joint RV+Hipparcos-IAD+Gaia-DR2/DR3 mass distribution, with Gaussian stellar-mass uncertainty ($1.156\pm0.031\,M_\odot$) propagated in post-processing. The red and dashed curves have unit peak; the solid curve and histogram share the product-density peak scale. Panel (d) is a diagnostic comparison with differing prior constructions; its solid curve is not a confidence envelope for the histogram.}
\label{fig:HD196068_RV}
\label{fig:HD196068_mass_constraint}
\end{figure*}

The white-noise+1p fit includes 36 AAT and 77 PFS radial velocities (Figure~\ref{fig:HD196068_RV}a--c). Its orbital summaries are $P={22.89}_{-0.82}^{+0.89}$~yr, $K={20.5}_{-1.3}^{+1.4}$~m\,s$^{-1}$, and $e={0.20}_{-0.04}^{+0.03}$. With the adopted stellar mass, the minimum companion mass is $M_{\min}={2.21}_{-0.15}^{+0.16}$~$M_{\rm Jup}$, and the semimajor axis evaluated at this minimum mass is ${8.47}_{-0.22}^{+0.23}$~au.

White noise+1p has a lower BIC than GP+1p by $1.4$ (Table~\ref{tab:RV_models}) and is adopted for HD~196068. Its instrumental and astrometric noise parameters are given in Table~\ref{tab:GP_hyperparameters}.

The joint RV, Hipparcos IAD, and Gaia posterior constrains the orbital period, eccentricity, and minimum mass, while the orbital orientation remains unconstrained. HD~196068~b is characterized as a long-period giant companion on a moderately eccentric orbit.

Panel~(d) of Figure~\ref{fig:HD196068_mass_constraint} presents a supplementary mass diagnostic from a separate white-noise+1p analysis, adapting the presentation of Figure~6 in \citet{Errico2022}. The astrometric constraint is computed from Hipparcos IAD and Gaia DR2/DR3 catalog measurements. Its product with the independent RV-only mass distribution has a similar main peak to the joint MCMC distribution, but the high-mass tails differ. The one-dimensional construction and the retained joint samples do not impose identical mass priors, so the product serves as an approximate diagnostic rather than an exact reconstruction of the joint mass posterior. We retain the minimum mass reported above as the adopted result.

\subsection{HD~203432}
\begin{figure*}[!ht]
\centering
\includegraphics[width=0.98\textwidth,height=0.72\textheight,keepaspectratio]{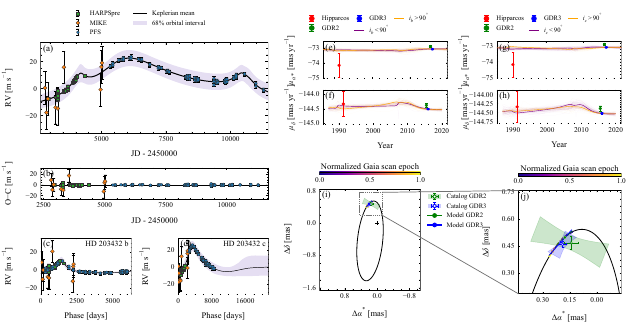}
\caption{Radial velocities and astrometric predictions for HD~203432 under the adopted GP+2p model. (a) RV time series. (b) Full-model residuals. (c,d) Phase-folded velocities of HD~203432~b and HD~203432~c after subtracting the companion signal. Instrument offsets and the chronological GP conditional posterior mean are subtracted from the RVs. Black curves show posterior-mean Keplerian signals; pale-purple shading encloses their pointwise 16th--84th percentile orbital intervals. RV error bars combine measurement errors and best-fit instrument jitter in quadrature. Residuals refer to the complete Keplerian+GP model. (e,f) Total proper motions in right ascension and declination, grouped by the inclination of HD~203432~b; (g,h) the same observables grouped by the inclination of HD~203432~c. Purple and orange curves show best-fit solutions within the two inclination hemispheres, and thin curves show 100 posterior predictions. Points denote Hipparcos and Gaia DR2/DR3 proper motions with formal errors. (i) Astrometric fitting for HD~203432~c; the dashed rectangle is enlarged in (j). The black curve shows the best-fit stellar reflex orbit. Open circles with error bars show conditional catalog positions; larger filled circles and short lines show model catalog projections. The contribution of HD~203432~b is subtracted at the same reference solution. Error bars retain formal position errors, while shading propagates catalog position and proper-motion errors. The colour scale marks normalized Gaia scan epoch.}
\label{fig:HD203432_RV}
\label{fig:HD203432_Ast_fit}
\end{figure*}

HD~203432 has a spectral classification of G8IV \citep{Gray2006} and is located at approximately $35.11$~pc. We use our stellar characterization in Table~\ref{tab:host_stars}, adopting $M_\star=1.04\pm0.04\,M_{\odot}$ and $T_{\rm eff}=5493\pm95$~K. The data comprise 74 HARPSpre, 9 MIKE, and 82 PFS radial velocities, spanning approximately $23.2$~yr in total. We jointly model the RV and Hipparcos--Gaia data with two Keplerian components and a GP contribution to the RV covariance (GP+2p). We denote the inner and outer companions as HD~203432~b and HD~203432~c, respectively. Figure~\ref{fig:HD203432_RV} shows the RV and astrometric fits.

For HD~203432, we compare white-noise+2p and GP+2p models using the same RV and astrometric data (Table~\ref{tab:HD203432_models}). GP+2p has the higher maximized joint likelihood and a lower BIC by $58.9$.

\begin{deluxetable*}{llrrrrr}[!ht]
\tabletypesize{\footnotesize}
\tablewidth{0pt}
\tablecaption{Joint-model comparison using maximum likelihood and BIC\label{tab:RV_models}\label{tab:HD203432_models}}
\tablehead{Target & Model & $k$ & $\ln\mathcal{L}_{\rm RV}(\hat{\theta})$ & $\ln\hat{\mathcal{L}}$ & BIC & $\Delta$BIC}
\startdata
HD~361 & White noise + 1p & $22$ & $-545.16$ & $-857.28$ & $1840.4$ & $97.7$ \\
HD~361 & GP + 1p & $24$ & $-490.59$ & $-802.72$ & $1742.7$ & $0.0$ \\
HD~196068 & White noise + 1p & $18$ & $-275.56$ & $-1297.09$ & $2696.7$ & $0.0$ \\
HD~196068 & GP + 1p & $20$ & $-270.58$ & $-1292.12$ & $2698.2$ & $1.4$ \\
HD~203432 & White noise + 2p & $27$ & $-420.08$ & $-709.67$ & $1572.9$ & $58.9$ \\
HD~203432 & GP + 2p & $29$ & $-388.01$ & $-674.51$ & $1513.9$ & $0.0$ \\
\enddata
\tablecomments{The normalized RV and joint log likelihoods are evaluated at the same maximum-likelihood solution. The joint likelihood includes RV, Hipparcos IAD, and Gaia catalogue terms. Instrumental RV offsets are profiled for this comparison and included in $k$. The observation counts are $N=305$, $298$, and $295$ for HD~361, HD~196068, and HD~203432, respectively. BIC and $\Delta$BIC follow Equation~\ref{eq:BIC}; differences are taken relative to the lowest BIC within each target.}
\end{deluxetable*}

The GP+2p posterior gives $A_{\rm GP}=3.21_{-0.30}^{+0.37}$~m\,s$^{-1}$ and $\tau_{\rm GP}=1.88_{-0.64}^{+0.65}$~d for the covariance in Equation~\ref{eq:gp_covariance}. Relative to the white-noise two-planet fit, the eccentricity medians decrease from approximately $0.74$ for both components to $0.63$ and $0.49$. The period median for HD~203432~c shifts from approximately $46227$ to $17049$~d, while the central period interval for HD~203432~b broadens from ${6250}_{-45}^{+60}$ to ${6335}_{-156}^{+175}$~d. Figure~\ref{fig:HD203432_RV} shows the GP-corrected velocities and Keplerian predictions; its residuals subtract the complete GP+2p model.

HD~203432~b has a period of ${6335}_{-156}^{+175}$~d (${17.35}_{-0.43}^{+0.48}$~yr), an RV semi-amplitude of ${7.6}_{-1.8}^{+2.2}$~m\,s$^{-1}$, and an eccentricity of ${0.63}_{-0.12}^{+0.11}$. Its semimajor axis is ${6.78}_{-0.15}^{+0.15}$~au and its minimum mass is $M_{\min,b}={0.55}_{-0.13}^{+0.15}$~$M_{\rm Jup}$.

HD~203432~c has a period of ${17049}_{-5122}^{+9831}$~d (${47}_{-14}^{+27}$~yr), an RV semi-amplitude of ${17.9}_{-3.5}^{+4.1}$~m\,s$^{-1}$, and an eccentricity of ${0.49}_{-0.13}^{+0.13}$. The corresponding semimajor axis is ${13.1}_{-2.8}^{+4.7}$~au and the minimum mass is $M_{\min,c}={2.03}_{-0.46}^{+0.55}$~$M_{\rm Jup}$.

The orbital posteriors include prograde and retrograde orientation solutions. For HD~203432~b, the conditional inclination summaries are ${62}_{-23}^{+19}$~deg and ${120}_{-21}^{+24}$~deg in the $i<90^\circ$ and $i>90^\circ$ hemispheres, respectively; the inclination and node distributions are broad. For HD~203432~c the corresponding values are ${60}_{-16}^{+18}$~deg and ${109}_{-14}^{+18}$~deg. Marginalizing over both hemispheres gives inclination-dependent mass summaries of ${0.66}_{-0.18}^{+0.32}$~$M_{\rm Jup}$ for HD~203432~b and ${2.43}_{-0.45}^{+0.48}$~$M_{\rm Jup}$ for HD~203432~c. These distributions include the stellar-mass propagation described above.

The period posterior for HD~203432~b also contains a concentration near $4200$~d, outside the central 16th--84th percentile interval. The marginal and cross-component posterior distributions are shown in Figure~\ref{fig:HD203432_corner_plot}.

\section{Discussion} \label{sec:discussion}

\subsection{HD~361}
HD~361~b resides on a wide, nearly circular orbit, with a semimajor axis of \(a = {7.35}_{-0.17}^{+0.17}\,\mathrm{au}\) and an eccentricity of \(e = {0.07}_{-0.03}^{+0.04}\). A plausible formation pathway involves the growth of a solid core in the outer protoplanetary disc, potentially facilitated by pebble accretion \citep{Lambrechts2012}, followed by the accretion of a gaseous envelope \citep{Pollack1996}. During the gas-rich phase, planet--disc interactions could have altered its semimajor axis through migration and damped its eccentricity, depending on the planet and disc properties \citep{Kley2012}.

The host star has a mildly subsolar metallicity of
\([\mathrm{Fe/H}] = -0.18 \pm 0.09\), making the solid inventory of the
primordial disc a relevant consideration in this formation scenario.
A sufficiently large reservoir of solids in the disc could have enabled the
planetary core to grow rapidly enough to initiate substantial gas accretion
before dispersal of the gaseous disc \citep{Ida2004}.

The existing RV data comprise periods of relatively dense temporal sampling interspersed with substantial observational gaps. 
Monitoring optimized for shorter orbital periods and sustained across multiple observing seasons would improve sensitivity to inner companions and constrain the architecture interior to HD~361~b.

\subsection{HD~196068}
HD~196068 forms a wide binary with HD~196067, whose giant planet was discovered by \citet{Marmier2013}. The projected stellar separation is approximately $664$~au, using the $16.62$~arcsec separation reported by \citet{Li2021} and our adopted parallax. The updated analysis of \citet{Delisle2025} places HD~196067~b at $a\simeq4.71$~au with $e\simeq0.56$ and identifies a low-mass inner planet with $P\simeq4.60$~d. HD~196068~b instead occupies a wider, moderately eccentric orbit at $a\simeq8.47$~au with $e\simeq0.20$. The presence of giant planets around both stars permits a comparison of their contrasting planetary architectures. HD~133131AB provides a comparable example, with giant planets orbiting both components \citep{Teske2016}. Core growth, potentially assisted by pebble accretion \citep{Lambrechts2012}, followed by gas accretion \citep{Pollack1996} and disc-driven evolution \citep{Kley2012}, remains a plausible formation pathway. Different initial disc conditions and migration histories could have contributed to the contrasting architectures. Tidal truncation \citep{Artymowicz1994} and forced planetesimal eccentricities \citep{Rafikov2015} provide possible ways for the stellar companion to influence their circumstellar discs.

After disc dispersal, the stellar companion may drive secular changes in the eccentricity and orbital orientation of HD~196068~b. An inclined binary--planet configuration could excite eccentricity through Kozai--Lidov oscillations \citep{Naoz2016}. The present eccentricity may therefore reflect a phase of long-term evolution, with its amplitude and timescale depending on the binary orbit and mutual inclination.

Epoch astrometry expected from Gaia DR4 will permit joint modelling of time-resolved astrometric measurements and RVs, improving the inclination constraint reported in Section~\ref{sec:results}. Continued RV monitoring and relative astrometry of the stellar pair would help distinguish the planetary reflex motion from the wide-binary contribution and constrain the binary orbit. Together, these measurements could improve the companion's inclination and true mass, determine its orientation relative to the binary, and test whether the geometries permitting strong secular excitation are compatible with the observed system.

\subsection{HD~203432}
\label{sec:discussion_hd203432}
The adopted GP+2p model describes an eccentric pair of cold giant planets, with the sub-Jovian HD~203432~b at approximately $6.8$~au and the more massive HD~203432~c at approximately $13.1$~au. Their minimum masses are approximately $0.55$ and $2.03\,M_{\rm Jup}$. The $17.35$~yr period of HD~203432~b is relatively well constrained, whereas the period of HD~203432~c, $47_{-14}^{+27}$~yr, exceeds the $23.2$~yr RV baseline. The observations constrain the signal of HD~203432~c through curvature over a partial orbit. A gap separates the HARPS and PFS observations, and MIKE coverage is sparse. Consequently, constraints from the observed curvature depend on both the inter-instrument velocity offsets and the orbital phase of HD~203432~c. The resulting asymmetric uncertainties are most pronounced in the period and semimajor axis of HD~203432~c and propagate into future orbital predictions. 

The RV model separates coherent orbital motion from variations correlated between nearby observations. Instrumental white-noise jitter accommodates excess scatter; the non-periodic Mat\'ern-$3/2$ covariance in Equation~\ref{eq:gp_covariance} accounts for correlations between observations. The inferred decay scale of approximately $1.9$~d is well separated from both orbital periods, supporting a decomposition into short-lived correlated variability and long-term Keplerian motion. Such correlations may arise from unresolved stellar variability or instrumental effects. The activity diagnostics in Appendix~\ref{app:stellar_activity} assess the stellar contribution independently. 

For the two-planet models, GP+2p has a lower BIC than white noise+2p by $58.9$ (Table~\ref{tab:HD203432_models}).

The eccentricities and modest orbital spacing make mutual perturbations relevant to the system's evolution. The semimajor axes differ by approximately a factor of two, and their quoted medians combined with the median eccentricities imply overlapping radial ranges. Secular coupling could drive apsidal precession and exchange eccentricity and inclination. Close approaches depend on the joint orbital geometry and phases, with the broad period distribution of HD~203432~c propagating into dynamical initial conditions. Improved mass and orientation constraints would support a stability analysis across the orbital posterior.

Extending the RV baseline will measure the evolution of the long-term curvature, separating period solutions for HD~203432~c and reducing their covariance with eccentricity and velocity offsets. Observations at epochs when competing orbital predictions for HD~203432~b diverge can test the period alternatives. Contemporaneous activity diagnostics and densely sampled RV sequences would constrain the short-timescale covariance and clarify its origin. Improved astrometry would further restrict the companion masses and mutual geometry, strengthening the initial conditions for dynamical calculations.

\section{Conclusions}
\label{sec:conclusions}
The joint RV and Hipparcos--Gaia analysis characterizes four long-period companions in three systems. HD~361~b has a low-eccentricity orbit at $7.35$~au and an inferred mass of $3.76\,M_{\rm Jup}$. HD~196068~b occupies a moderately eccentric orbit at $8.47$~au, with a minimum mass of $2.21\,M_{\rm Jup}$; its inclination remains unconstrained. HD~203432 is described by an eccentric pair at approximately $6.8$ and $13.1$~au. The GP+2p model has a lower BIC than white noise+2p. Astrometry informs the companion masses and orbital orientations, with prograde and retrograde solutions retained. These systems span distinct cold-giant architectures, including a nearly circular giant, a companion in a wide stellar binary, and an eccentric two-planet system. Continued RV monitoring will refine the long-period orbits and search for additional inner companions. Time-resolved astrometry will improve mass and orientation constraints, providing a firmer basis for investigating their dynamical evolution.

%% Please use the acknowledgment and contribution environments. This will 
%% be anonomyized when the "anonymous" style option is used. 

\begin{acknowledgments}
This work is supported by the National Key R\&D Program of China, No.~2024YFA1611801 and No.~2024YFC2207700, by the National Natural Science Foundation of China (NSFC) under Grant No.~12473066, by the Shanghai Jiao Tong University 2030 Initiative, and by the China-Chile Joint Research Fund (CCJRF No. 2205). CCJRF is provided by the Chinese Academy of Sciences South America Center for Astronomy (CASSACA) and established by the National Astronomical Observatories, Chinese Academy of Sciences (NAOC), and Chilean Astronomy Society (SOCHIAS) to support China-Chile collaborations in astronomy. The authors acknowledge the years of technical support from LCO staff in the successful operation of PFS, enabling the collection of the data presented in this paper. 
The computations in this paper were run on the $\pi$ 2.0 (or the Siyuan-1) cluster supported by the Center for High Performance Computing at Shanghai Jiao Tong University. This work is partly based on observations collected at the European Organisation for Astronomical Research in the Southern Hemisphere under ESO programs: 108.22KV, 60.A-9036, 072.C-0488, 079.C-0681, 087.D-0511, 091.C-0844, 094.C-0894, 192.C-0852, 196.C-1006, 105.20MP.001, 108.2271.002 and 108.2271.003. Part of this research was carried out at the Jet Propulsion Laboratory, California Institute of Technology, under a contract with the National Aeronautics and Space Administration (80NM0018D0004).
We thank the developers and maintainers of \texttt{Anaconda}, \texttt{Jupyter Notebook}, and \texttt{Spyder} for their software.
\end{acknowledgments}

\facilities{Magellan:Clay (PFS, MIKE), ESO:3.6m (HARPS), AAT (UCLES), Hipparcos, Gaia}

\software{Astropy \citep{Astropy2013,Astropy2018,Astropy2022}, NumPy \citep{harris2020array}, SciPy \citep{virtanen2020scipy}, pandas \citep{2020zndo...3898987R}, Matplotlib \citep{hunter2007matplotlib}, seaborn \citep{waskom2021seaborn}, corner \citep{foreman2016corner}, ptemcee \citep{Vousden2016}}


\appendix
\section{Stellar activity diagnostics}
\label{app:stellar_activity}
\begin{figure*}[!ht]
\centering
\IfFileExists{Activity_GLS_and_Pearson_adopted.pdf}{
\includegraphics[width=0.96\textwidth,height=0.68\textheight,keepaspectratio]{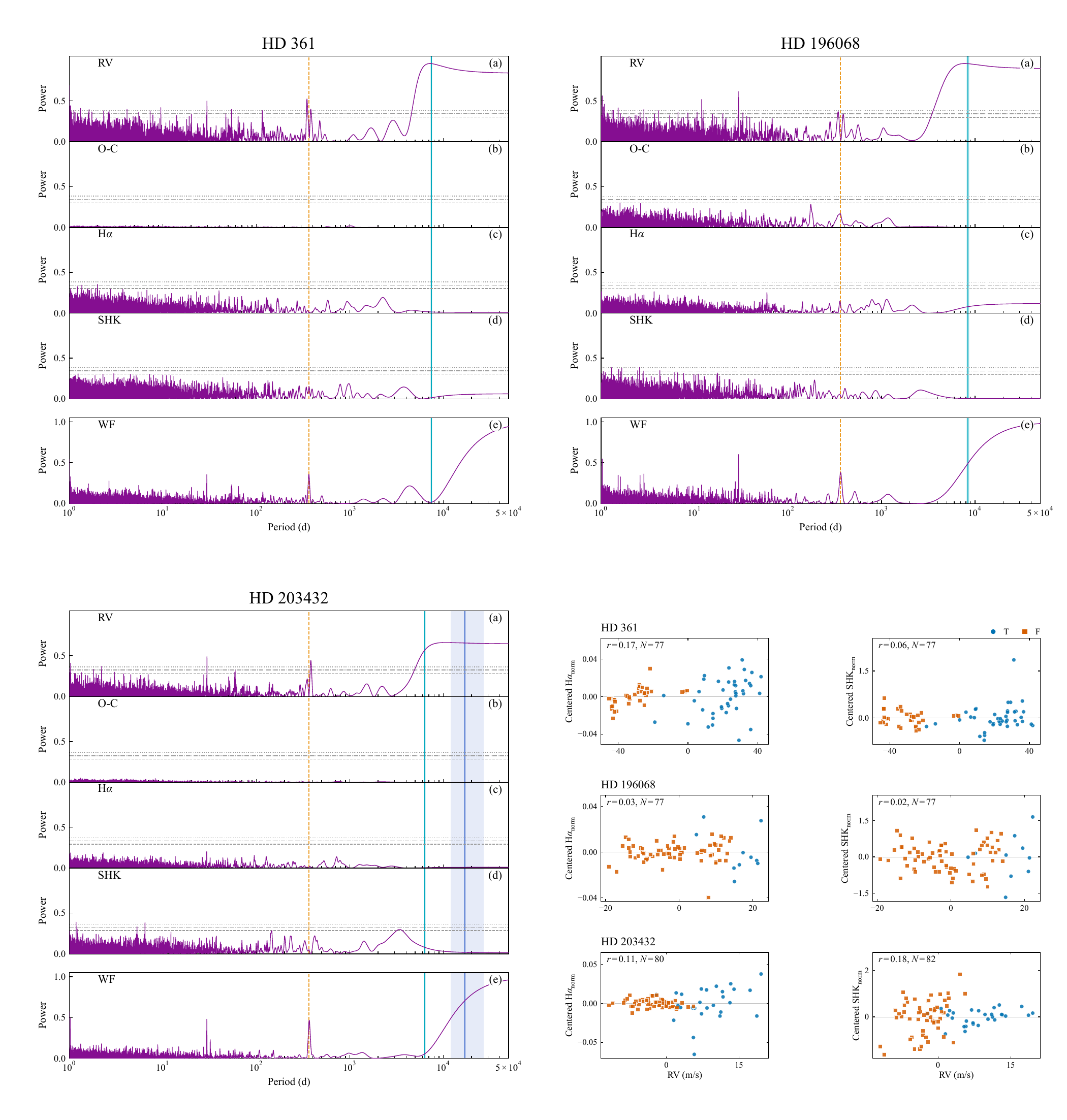}
}{
\fbox{\parbox[c][0.18\textheight][c]{0.90\textwidth}{\centering Figure awaiting upload\\[1ex]
\texttt{\detokenize{Activity_GLS_and_Pearson_adopted.pdf}}}}
}
\caption{
PFS RV and stellar-activity diagnostics for HD~361 (upper left),
HD~196068 (upper right), and HD~203432 (lower left).
Each block shows GLS periodograms of (a) RV, (b) full-model residuals
(O$-$C), (c) H$\alpha$, and (d) SHK, followed by (e) the RV sampling
window (WF).
T and F label the two calibration-spectrum groups in the PFS activity tables.
Activity indices are divided by their respective group medians,
and each GLS fits one floating mean.
O$-$C subtracts the adopted Keplerian posterior-mean predictions and
instrumental offsets. For HD~361 and HD~203432, the GP conditional
posterior mean at the observing epochs is also subtracted.
Solid vertical lines and shaded bands mark the companion periods and
their 16th--84th percentile intervals; orange dashed lines mark 365.25~d.
Grey dashed, dash-dotted, and dotted lines indicate nominal white-noise
FAP levels of 1\%, 0.1\%, and 0.01\%, respectively.
The lower-right block shows Pearson correlations for the three stars
in successive rows, with H$\alpha$ on the left and SHK on the right.
For these panels only, the normalized indices are additionally centered
by their T/F group means; RVs remain unchanged.
Colours distinguish T/F groups, and annotations give the unweighted
Pearson coefficient $r$ and sample size $N$.
}
\label{fig:activity_diagnostics}
\end{figure*}

Figure~\ref{fig:activity_diagnostics} compares PFS RVs, activity periodograms, the sampling window, and RV--activity correlations for the three systems. The GLS periodograms show no significant peaks near the modeled companion periods in either the $S_{\text{HK}}$ or $\mathrm{H}\alpha$ time series for any of the three systems. All three stars exhibit relatively quiet activity signals, and the RVs show no significant Pearson correlation with either activity indicator. The full-model residual periodograms also lie below the plotted nominal 1\% white-noise false-alarm reference level. These results suggest that stellar activity is unlikely to be the source of the planet signals.

\clearpage
\section{Posterior distributions}
\label{app:posteriors}
\begin{figure*}[!ht]
\centering
\IfFileExists{HD361_orbital_GP_corner.pdf}{
\includegraphics[width=0.96\textwidth,height=0.68\textheight,keepaspectratio]{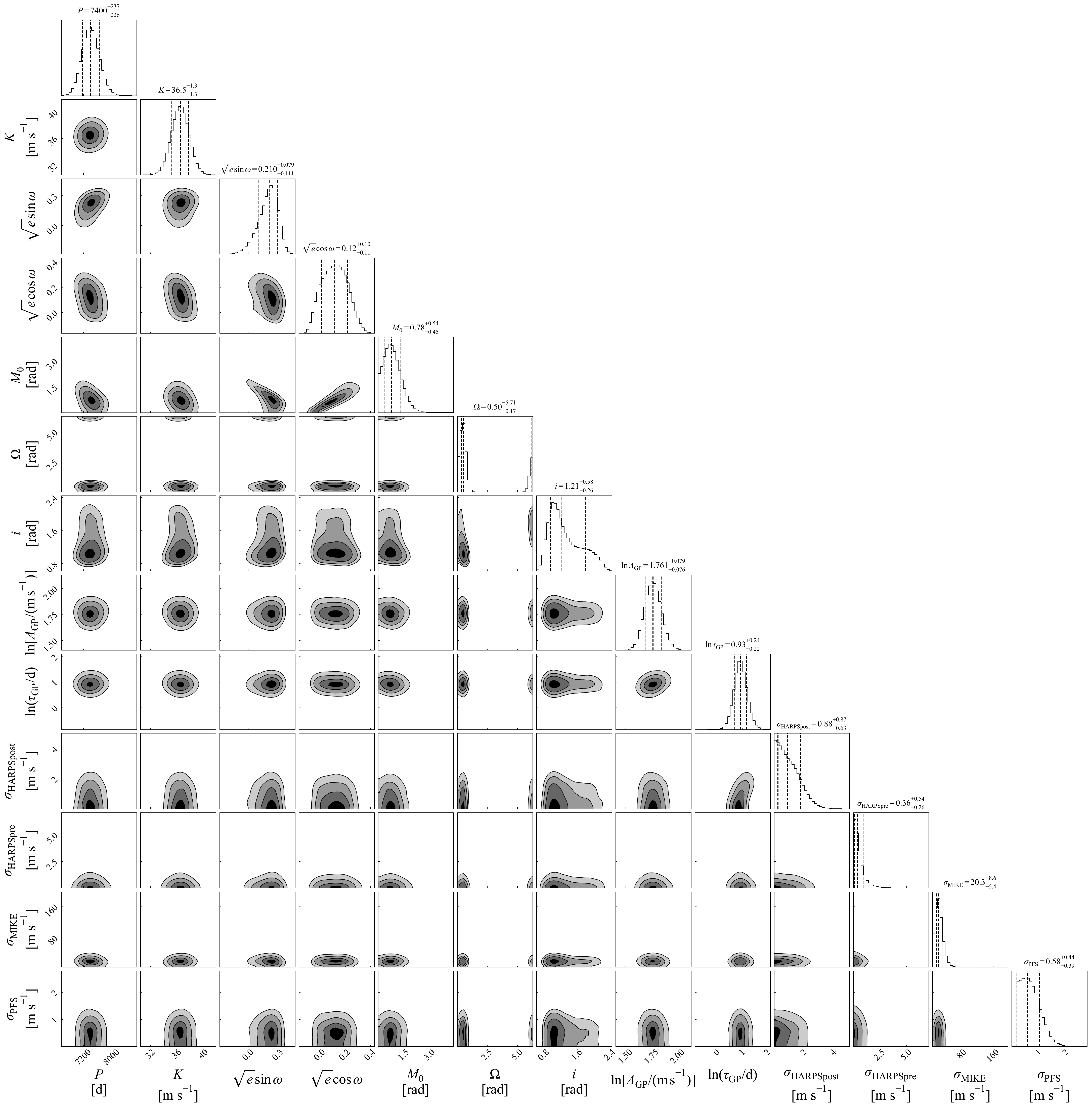}
}{
\fbox{\parbox[c][0.18\textheight][c]{0.90\textwidth}{\centering Figure awaiting upload\\[1ex]
\texttt{\detokenize{HD361_orbital_GP_corner.pdf}}}}
}
\caption{Fitted parameter distributions for HD~361 under GP+1p, from 80000 posterior samples. The lower-left triangle contains the orbital parameters, GP hyperparameters, and RV jitters. Dashed lines mark the 16th, 50th, and 84th percentiles; annotations give the median and equal-tail uncertainties. Shading and contours show two-dimensional marginalized densities. The GP amplitude and decay time are shown in their fitted natural-log coordinates, with amplitude in m\,s$^{-1}$ and time in days. The mean anomaly $M_0$ refers to J2017.5. Angular coordinates are linear, with units specified on the axes.}
\label{fig:HD361_corner_plot}
\end{figure*}

\begin{figure*}[!ht]
\centering
\IfFileExists{HD196068_RV_derived_WN_corner.pdf}{
\includegraphics[width=0.96\textwidth,height=0.68\textheight,keepaspectratio]{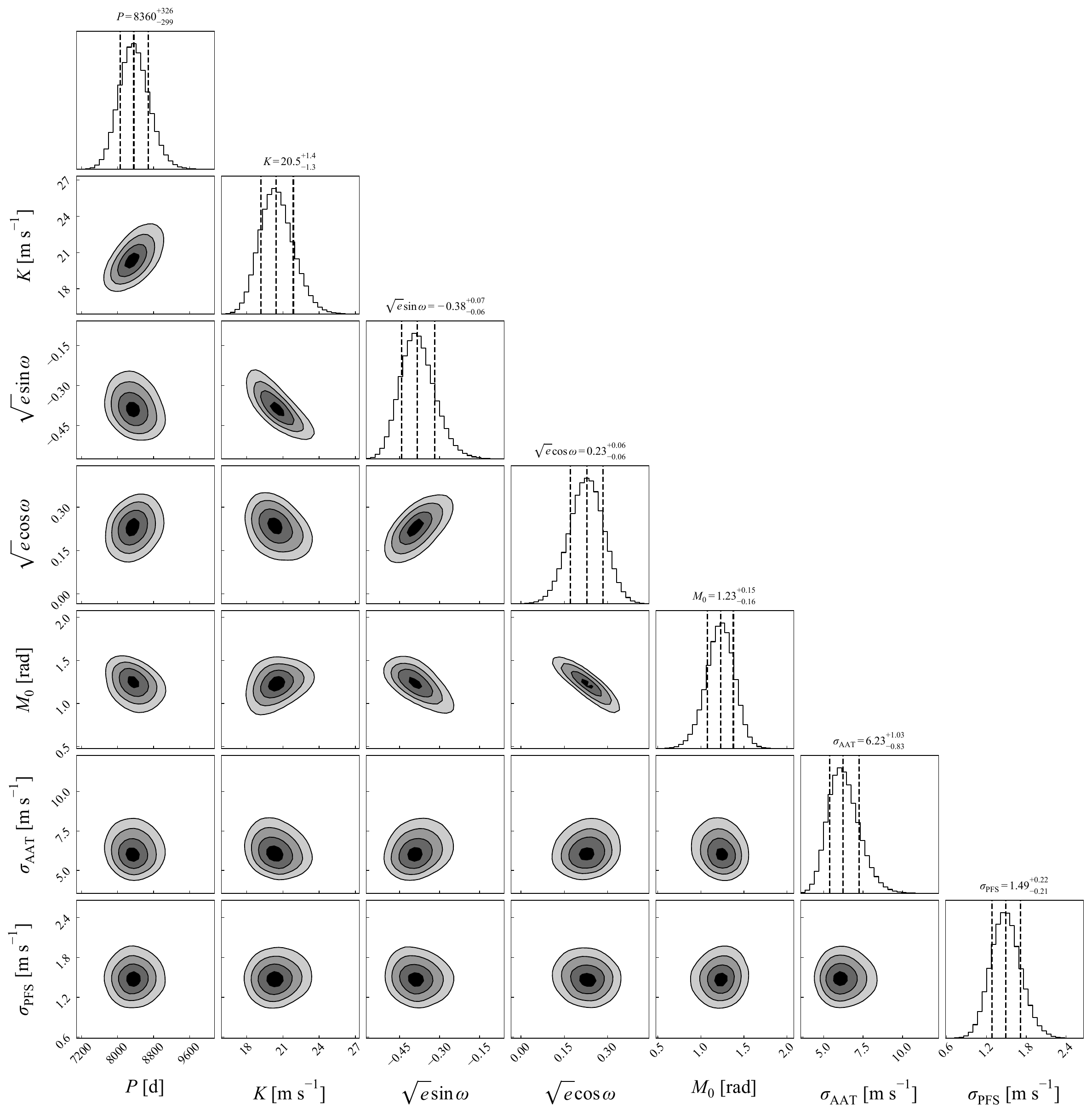}
}{
\fbox{\parbox[c][0.18\textheight][c]{0.90\textwidth}{\centering Figure awaiting upload\\[1ex]
\texttt{\detokenize{HD196068_RV_derived_WN_corner.pdf}}}}
}
\caption{Fitted RV parameter distributions for HD~196068 under white noise+1p, from 80000 joint-posterior samples. The lower-left triangle contains the orbital parameters and AAT and PFS jitters. The mean anomaly refers to J2017.5; percentile and contour conventions follow Figure~\ref{fig:HD361_corner_plot}.}
\label{fig:HD196068_corner_plot}
\end{figure*}

\begin{figure*}[!ht]
\centering
\IfFileExists{HD203432_joint_GP_corner.pdf}{
\includegraphics[width=0.96\textwidth,height=0.68\textheight,keepaspectratio]{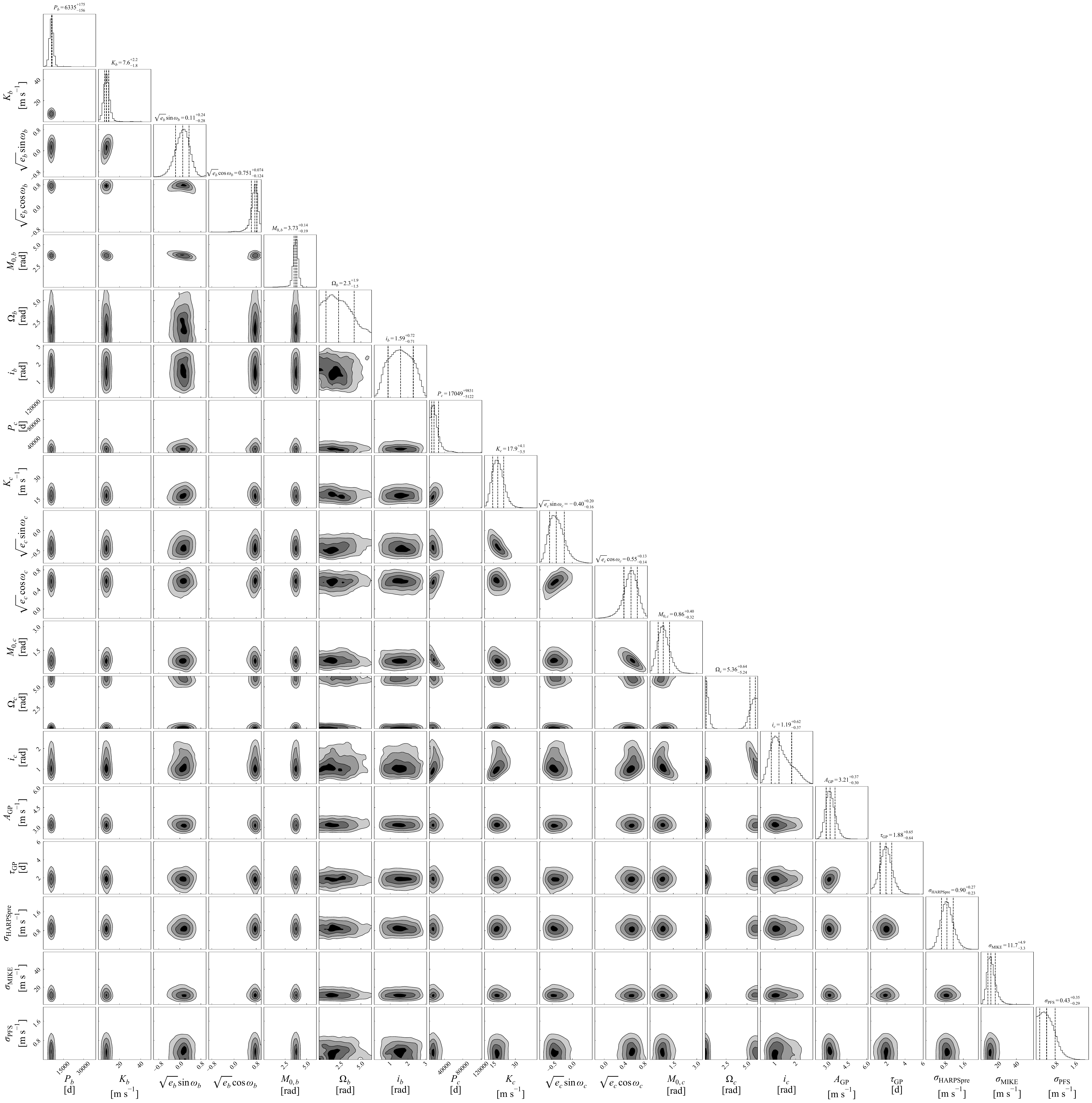}
}{
\fbox{\parbox[c][0.18\textheight][c]{0.90\textwidth}{\centering Figure awaiting upload\\[1ex]
\texttt{\detokenize{HD203432_joint_GP_corner.pdf}}}}
}
\caption{Joint fitted parameter distributions for HD~203432~b and HD~203432~c under the GP+2p model, from 80000 posterior samples including alternative period and orientation solutions. The lower-left triangle contains both sets of orbital parameters, the GP amplitude and decay timescale, and the instrumental RV jitters. The mean anomalies refer to J2017.5. Dashed lines mark the 16th, 50th, and 84th percentiles; annotations give the medians and equal-tail uncertainties. Contours show two-dimensional marginalized densities. Angular coordinates are linear, with units specified on the axes.}
\label{fig:HD203432_corner_plot}
\end{figure*}

Figures~\ref{fig:HD361_corner_plot}--\ref{fig:HD203432_corner_plot} show the marginal and joint orbital posterior distributions, including the alternative period and orientation solutions described in Section~\ref{sec:results}.

%% For this sample we use BibTeX plus aasjournalv7.bst to generate the

\clearpage
\bibliography{sample701}{}
\bibliographystyle{aasjournalv7.1}



\end{document}